\documentclass
[aps,pre,twocolumn,groupedaddress,showpacs,preprintnumbers,floatfix]{revtex4}

\usepackage[tbtags]{amsmath}
\usepackage{bm,graphicx}
\DeclareMathOperator{\tr}{tr}
\newcommand{\di}[1]{\mathsf{#1}}
\newcommand{\uc}[0]{u_{\text{c}}}
\newcommand{\zc}[0]{z_{\text{c}}}
\begin{document}

\preprint{SAND2007-8046J}

\title{Fronts in randomly advected and heterogeneous media and\\
nonuniversality of Burgers turbulence: Theory and numerics}

\author{Jackson R. Mayo}
\email{jmayo@sandia.gov}
\author{Alan R. Kerstein}
\email{arkerst@sandia.gov}
\affiliation{Combustion Research Facility, Sandia National Laboratories,
Livermore, California 94551-0969, USA}

\begin{abstract}
A recently established mathematical equivalence---between weakly perturbed
Huygens fronts (e.g., flames in weak turbulence or geometrical-optics wave
fronts in slightly nonuniform media) and the inviscid limit of
white-noise-driven Burgers turbulence---motivates theoretical and numerical
estimates of Burgers-turbulence properties for specific types of white-in-time
forcing. Existing mathematical relations between Burgers turbulence and the
statistical mechanics of directed polymers, allowing use of the replica method,
are exploited to obtain systematic upper bounds on the Burgers energy density,
corresponding to the ground-state binding energy of the directed polymer and
the speedup of the Huygens front. The results are complementary to previous
studies of both Burgers turbulence and directed polymers, which have focused on
universal scaling properties instead of forcing-dependent parameters. The
upper-bound formula can be heuristically understood in terms of renormalization
of a different kind from that previously used in combustion models, and also
shows that the burning velocity of an idealized turbulent flame does not
diverge with increasing Reynolds number at fixed turbulence intensity, a
conclusion that applies even to strong turbulence. Numerical simulations of the
one-dimensional inviscid Burgers equation using a Lagrangian finite-element
method confirm that the theoretical upper bounds are sharp within about 15\%
for various forcing spectra (corresponding to various two-dimensional random
media). These computations provide a new quantitative test of the replica
method. The inferred nonuniversality (spectrum dependence) of the front speedup
is of direct importance for combustion modeling.
\end{abstract}


\maketitle

\section{\label{Intro}Introduction}

In the geometrical optics of a ``quenched'' medium with static spatial
variations in refractive index, or in the combustion of a heterogeneous solid
propellant, each spatial point is characterized by a local speed
$v(\mathbf{x})$ at which a wave front or flame sheet can advance normal to
itself in accordance with Huygens' principle \cite{KA94}. A key problem is to
determine the time needed for an influence (light or combustion) to propagate
from one region of space to another; Fermat's principle asserts that the
propagation effectively occurs along all possible trajectories that obey the
local speed $v(\mathbf{x})$, with the first-arriving trajectory giving the
desired answer.

A preliminary connection with statistical mechanics is suggested by writing the
travel time for a spatial path $C$ as
\begin{equation}
\label{tC}
t(C) = \int_C ds\, \frac{1}{v(\mathbf{x})} = \langle v^{-1}\rangle\, l(C) +
\int_C ds\, \sigma(\mathbf{x}),
\end{equation}
where $l(C)$ is the path length, $\langle v^{-1}\rangle$ is a spatial average
over the medium, and $\sigma = v^{-1} - \langle v^{-1}\rangle$ is a zero-mean
field. We can interpret $C$ as a material curve---a polymer string---and $t(C)$
as its potential energy, with $\langle v^{-1}\rangle$ being the constant
tension and $\sigma$ an external potential. The polymer is infinitely flexible,
since there is no curvature energy in Eq.\ (\ref{tC}). And the polymer is
``directed'' because each endpoint is confined to a different region of space.
Fermat's principle requires us to find the minimum-energy configuration
(classical ground state) of this directed polymer.

So far this is just a complicated optimization problem, but it can be made more
explicitly statistical in two ways. First, in many cases of interest, $\sigma$
is a homogeneous random field with specified statistics, and we are satisfied
with calculating the ensemble-averaged propagation time or minimum energy
(which, for a very long polymer, nearly equals the minimum energy in a single
realization of $\sigma$). Second, even within a given realization of the
potential, it may be convenient to consider the canonical ensemble of
directed-polymer configurations at finite temperature, calculate the
thermodynamic free energy, and then take the zero-temperature limit to obtain
the ground-state energy. The most powerful theoretical methods arise from
combining both types of ensemble.

This paper focuses on the case in which $\sigma$ is a homogeneous and isotropic
(but not necessarily Gaussian) random field that is small in magnitude compared
to $\langle v^{-1}\rangle$. In this weak-fluctuation limit, there is in fact a
more general and useful relation between propagation and directed polymers, one
that applies even to random media with time dependence and advection, such as
turbulent fluids. In this paper, we focus on the quenched-medium interpretation
of the analysis, but relate our approach to turbulent combustion and other
applications when useful. As was recently shown \cite{MK07}, the propagation of
Huygens fronts in a broad class of weakly random $d$-dimensional media reduces
to the $(d - 1)$-dimensional inviscid Burgers equation with white-in-time
forcing. Furthermore, the statistical properties of the inviscid Burgers
dynamics are equivalent to those of viscous ``Burgers turbulence'' (with the
same forcing) in the limit of vanishing viscosity $\nu$ \cite{GIKP05}. The
white-noise-driven viscous Burgers equation, in turn, has a known relation to
the canonical ensemble for a near-straight directed polymer (at a temperature
proportional to $\nu$) in a $d$-dimensional static random potential that is
white in one direction \cite{BMP95}. This is precisely the kind of random
potential commonly assumed in theoretical directed-polymer studies
\cite{MP91,G93,B94}.

In Sec.~\ref{Relations} we systematically discuss some known relations among
different theoretical representations of our problem, and indicate how the
recent results can be linked with this framework to allow calculation of the
speedup of Huygens fronts. In Sec.~\ref{Monotonicity} we describe three
rigorous ``monotonicity properties'' obeyed by random Huygens propagation,
which provide not only checks on our later results but also further
implications from them. In Sec.~\ref{Replica} we present the replica method in
a form suited to our problem, and derive explicit upper bounds on the front
speedup. In Sec.~\ref{Numerical} we review existing numerical simulations that
provide relevant speedup values for comparison, and describe our new
higher-precision simulation results. A summary and discussion are presented in
Sec.~\ref{Discussion}.

\section{\label{Relations}Relations among theories}

\subsection{\label{KPZBH}KPZ, Burgers, and Huygens}

The Kardar-Parisi-Zhang (KPZ) equation for interface growth \cite{KPZ86} is
well known as a unifying model for diverse phenomena. The equation can be
written
\begin{equation}
\label{KPZ}
\frac{\partial h}{\partial t} = \tfrac{1}{2} |\bm{\nabla}_\perp h|^2 +
\nu\nabla_\perp^2 h + \eta(t,\mathbf{x}_\perp),
\end{equation}
where $h(t,\mathbf{x}_\perp)$ is a fluctuation in the height of an interface as
a function of $d - 1$ transverse dimensions and time, $\nu > 0$ is a surface
tension that smooths the interface, and $\eta(t,\mathbf{x}_\perp)$ is a
zero-mean external perturbation (not necessarily the white noise assumed by
KPZ). Equation (\ref{KPZ}) is equivalent to the $(d - 1)$-dimensional forced
viscous Burgers equation
\begin{equation}
\label{Burgers}
\frac{\partial\mathbf{w}}{\partial t} + (\mathbf{w} \cdot \bm{\nabla}_\perp)
\mathbf{w} = \nu\nabla_\perp^2 \mathbf{w} - \bm{\nabla}_\perp \eta
\end{equation}
for the velocity field $\mathbf{w} = -\bm{\nabla}_\perp h$, and also to the $(d
- 1)$-dimensional imaginary-time Schr\"odinger equation for the ``wave
function'' $\exp(h/2\nu)$, with potential $-\eta$ and Planck's constant $2\nu$.
The solution of this Schr\"odinger equation is given by the Feynman path
integral \cite{FH65}
\begin{multline}
\label{Feyn}
\exp\frac{h(t,\mathbf{x}_\perp)}{2\nu} = \int_{\mathbf{y}(t) =
\mathbf{x}_\perp} \mathcal{D}\mathbf{y}(u)\,
\exp\biggl(\frac{h\bm{(}0,\mathbf{y}(0)\bm{)}}{2\nu}\\
- \frac{1}{2\nu} \int_0^t du\, [\tfrac{1}{2} |\mathbf{y}'(u)|^2 -
\eta\bm{(}u,\mathbf{y}(u)\bm{)}]\biggr).
\end{multline}

The simplest connection between these ideas and Huygens propagation is to
assume a quenched medium with weak refractive-index fluctuations given by a
smooth function $\sigma(x_\parallel,\mathbf{x}_\perp)$, and to write the
position of a nearly flat front as the unperturbed position plus a small
displacement:
\begin{equation}
\label{xth}
x_\parallel(t,\mathbf{x}_\perp) = t + h(t,\mathbf{x}_\perp),
\end{equation}
in units where $\langle v^{-1}\rangle = 1$. We can approximate the incremental
speed $\eta(t,\mathbf{x}_\perp)$ in Eq.\ (\ref{KPZ}) as the negative of the
index fluctuation at the unperturbed location, i.e., $\eta(t,\mathbf{x}_\perp)
\simeq -\sigma(t,\mathbf{x}_\perp)$. Equation (\ref{KPZ}) then governs the
dynamics of $h$, with the nonlinear term describing the leading-order effect of
tilted propagation, and the Laplacian term smoothing the cusps (or Burgers
shocks) that would otherwise develop. (We work to leading order in the
fluctuation $\sigma$ and the tilt $\bm{\nabla}_\perp h$, neglecting their
higher-order and joint effects.) The parameter $\nu$, which corresponds to the
Markstein length in premixed flamelet combustion \cite{W85}, contributes a
stabilizing dependence of front speed on local curvature, with concave regions
propagating faster than convex regions. Pure Huygens propagation is obtained as
a ``viscosity solution'' in the $\nu \to 0$ limit \cite{CL83,S85}.

A useful alternative description is based on a field
$T_0(x_\parallel,\mathbf{x}_\perp)$ giving the time at which the front reaches
a given point. From Eq.\ (\ref{xth}), we see that to leading order
\begin{equation}
\label{Txh}
T_0(x_\parallel,\mathbf{x}_\perp) = x_\parallel -
h(x_\parallel,\mathbf{x}_\perp).
\end{equation}
Thus, in the weak-fluctuation framework, we are free to write $x_\parallel$ in
place of $t$ as an independent variable and interpret $-h$ as the time
deviation. With this perspective, the path integral (\ref{Feyn}) is, upon
suitable normalization, the canonical partition function for the simple
``Fermat's principle'' directed polymer described in the Introduction. With
weak fluctuations, only paths that are nearly aligned with the
$x_\parallel$-direction are competitive in travel time (energy). Such a path,
described by a function $\mathbf{y}(u)$ so that $\mathbf{x}_\perp =
\mathbf{y}(x_\parallel)$, has arc-length element $ds = du\, [1 + \tfrac{1}{2}
|\mathbf{y}'(u)|^2]$ and travel time
\begin{equation}
\begin{split}
t\bm{(}\mathbf{y}(u)\bm{)} &= \int ds\, [1 +
\sigma\bm{(}u,\mathbf{y}(u)\bm{)}]\\
&= \int du\, [1 + \tfrac{1}{2} |\mathbf{y}'(u)|^2 +
\sigma\bm{(}u,\mathbf{y}(u)\bm{)}].
\end{split}
\end{equation}
Then, assuming a flat initial front with $h = 0$, Eq.\ (\ref{Feyn}) asserts
that
\begin{widetext}
\begin{align}
\exp\frac{-T_0(x_\parallel,\mathbf{x}_\perp)}{2\nu} &=
\exp\frac{-x_\parallel}{2\nu} \times \int_{\mathbf{y}(x_\parallel) =
\mathbf{x}_\perp} \mathcal{D}\mathbf{y}(u)\, \exp\biggl(-\frac{1}{2\nu}
\int_0^{x_\parallel} du\, [\tfrac{1}{2} |\mathbf{y}'(u)|^2 +
\sigma\bm{(}u,\mathbf{y}(u)\bm{)}]\biggr)\nonumber\\
\label{partf}
&= \int_{\mathbf{y}(x_\parallel) = \mathbf{x}_\perp} \mathcal{D}\mathbf{y}(u)\,
\exp\frac{-t\bm{(}\mathbf{y}(u)\bm{)}}{2\nu}.
\end{align}
\end{widetext}
With the travel time as the energy, the right-hand side is the partition
function at temperature $2\nu$ for a polymer with one end constrained to the
initial front and the other end fixed at $(x_\parallel,\mathbf{x}_\perp)$.
Consequently, $T_0(x_\parallel,\mathbf{x}_\perp)$ is the free energy of this
polymer. As expected, Eq.\ (\ref{partf}) indicates that $T_0$ approaches the
absolute minimum travel time (ground-state energy) in the Huygens-propagation
limit $\nu \to 0$.

If $\sigma$ is a homogeneous random field, we expect the free energy $T_0$ to
scale linearly with the longitudinal extent $x_\parallel$ of the polymer in the
thermodynamic limit $x_\parallel \to \infty$. Since the ground-state energy in
the absence of fluctuations would be just $x_\parallel$, we define the binding
energy per unit length as
\begin{equation}
\Delta = \lim_{x_\parallel\to\infty} \left(1 -
\frac{T_0(x_\parallel,\mathbf{x}_\perp)}{x_\parallel}\right) =
\lim_{x_\parallel\to\infty} \frac{h(x_\parallel,\mathbf{x}_\perp)}{x_\parallel}
> 0,
\end{equation}
which is independent of $\mathbf{x}_\perp$ by homogeneity, and is positive
because the nonlinear term in Eq.\ (\ref{KPZ}) makes $h$ increase on average.
This nonlinear term equals $\tfrac{1}{2} w^2$, and so $\Delta$ can also be
described as the steady-state energy density of the Burgers fluid. [The other
two terms on the right-hand side of Eq.\ (\ref{KPZ}) average to zero because
$h$ is statistically homogeneous in $\mathbf{x}_\perp$ and $\eta$ is a centered
perturbation.] The effect of the weak fluctuations $\sigma$, whose rms value we
denote by $\epsilon \ll 1$, is to renormalize the overall front speed upward to
$v_* = \lim_{x_\parallel\to\infty} x_\parallel/T_0 = 1 + \Delta$. Our analysis
applies to the asymptotic limit $\epsilon \to 0$ and is expected to be accurate
when the dimensionless parameter $\epsilon$ is small; previous numerical work
\cite{RMS93} discussed in Sec.~\ref{Existing} indicates that the asymptotic
scaling holds within a few percent for $\epsilon \le 0.1$.

\subsection{\label{White}White-noise reduction}

For media of a given structure, with fluctuations related by overall rescaling,
it is clear that $\Delta \to 0$ as $\epsilon \to 0$; but the form of this
dependence is subtle. If $\nu$ is fixed, then for sufficiently small
$\epsilon$, Eq.\ (\ref{KPZ}) can be linearized; the lowest-order solution for
$h$ is proportional to $\epsilon$ and thus the nonlinear term (producing the
secular growth of $h$ measured by $\Delta$) is proportional to $\epsilon^2$.
This is the weak-perturbation scenario normally associated with the KPZ
equation, corresponding to a laminar solution of the viscous Burgers equation
(\ref{Burgers}). On the other hand, if we take $\nu \to 0$ at finite $\epsilon$
to obtain Huygens propagation, and only then take $\epsilon \to 0$, the
linearization is invalid (because the Burgers flow is fully turbulent) and the
scaling of $\Delta$ with $\epsilon$ is not immediately obvious. Subject to mild
conditions on the medium structure, we have shown mathematically \cite{MK07}
that $\Delta \propto \epsilon^{4/3}$ in this regime, confirming a previous
conjecture \cite{KA92}. Here we provide a somewhat more physical argument based
on dimensional analysis.

Let us denote by $\alpha_\parallel$ and $\alpha_\perp$ some measures of the
longitudinal and lateral correlation lengths of $\sigma$, and consider the
behavior of the inviscid Burgers equation upon varying these quantities and
$\epsilon$, while keeping fixed all other details of the statistics of
$\sigma$. We can perform dimensional analysis on the KPZ equation (\ref{KPZ}),
with $\eta = -\sigma$ and $\nu = 0$, by assigning the conventional dimensions
from the Burgers-fluid interpretation:
\begin{gather}
[w] = [\nabla_\perp h] = \di L \di T^{-1}, \quad [h] = \di L^2 \di T^{-1},
\quad [\alpha_\parallel] = \di T,\nonumber\\
[\alpha_\perp] = \di L, \quad [\eta] = [\epsilon] = \di L^2 \di T^{-2}, \quad
[\Delta] = \di L^2 \di T^{-2}.
\end{gather}
(These dimensions are unusual from the viewpoint of propagation because the
longitudinal and lateral directions are treated differently.) The most general
dimensionless combination of input parameters is a function of $q =
\alpha_\parallel^2 \alpha_\perp^{-2} \epsilon$. Accordingly,
\begin{equation}
\label{dim}
\Delta = (\alpha_\parallel \alpha_\perp^{-1} \epsilon^2)^{2/3}\, Q(q),
\end{equation}
where we have taken advantage of the freedom to choose any combination of
parameters with the dimensions of $\Delta$ to multiply $Q(q)$. (We could
equally well have chosen $\epsilon$ or $\alpha_\parallel^{-2} \alpha_\perp^2$.)

The reason for our choice is that if we take $\alpha_\parallel \to 0$ while
scaling $\epsilon \propto \alpha_\parallel^{-1/2}$, so that $\sigma$ becomes
white noise in the longitudinal (time) direction, then the coefficient of
$Q(q)$ is unaffected. Assuming that the inviscid Burgers equation is well
behaved with white-in-time forcing, so that $\Delta$ remains finite, we can
partially constrain the function $Q$. The fact that $\epsilon \to \infty$ as
$\alpha_\parallel \to 0$ is of no concern here, because the sole formally
dimensionless measure of the strength of $\epsilon$ is $q$, which is
approaching zero. We conclude that $Q(q)$ has a finite limit as $q \to 0$. Now,
keeping the correlation lengths fixed but taking $\epsilon \to 0$ (which also
gives $q \to 0$), we see that $\Delta \propto \epsilon^{4/3}$.

Our mathematical analysis \cite{MK07} demonstrated this result more seamlessly,
starting from the exact equations of Huygens propagation (rather than from the
KPZ equation), extracting a factor $\epsilon^{4/3}$ by a rescaling of
variables, and directly obtaining the white-noise-driven inviscid Burgers
equation as $\epsilon \to 0$. (To make the argument used in this paper more
complete, in Appendix \ref{appKPZ} we provide a mathematical justification for
starting from the KPZ equation.) Furthermore, we showed that the effect of an
advecting velocity field $\mathbf{u}$ of order $\epsilon$---along with possible
time dependence of $\sigma$ and $\mathbf{u}$ on a natural timescale of order
$\epsilon^{-1}$---is given as $\epsilon \to 0$ by simply adding $-u_\parallel$
to $\sigma$ and considering the medium as quenched at the initial time. That
is, both the component $\mathbf{u}_\perp$ and the time dependence of the medium
become irrelevant.

If we reintroduce the viscosity (Markstein length) $\nu$, with conventional
dimensions $\di L^2 \di T^{-1}$, there is a new dimensionless parameter that
remains fixed in the white-noise limit: the Burgers-fluid Reynolds number
$\mathrm{Re_B} \sim (\alpha_\parallel \alpha_\perp^2 \epsilon^2)^{1/3}
\nu^{-1}$. To maintain Huygens propagation (corresponding to very large
$\mathrm{Re_B}$) as $\epsilon \to 0$, we must decrease $\nu$ at least as fast
as $\epsilon^{2/3}$. In our previous analysis, we cited mathematical results
\cite{IK03,GIKP05} establishing that the white-noise-driven steady state of the
inviscid Burgers equation exists, and is approached by that of the viscous
Burgers equation as $\mathrm{Re_B} \to \infty$. Thus there is no subtlety with
interchange of the limits $\epsilon \to 0$ and $\mathrm{Re_B} \to \infty$, as
there was with $\epsilon \to 0$ and $\nu \to 0$ in Eq.\ (\ref{KPZ}). Previously
\cite{MK07} we denoted by $\nu$ what we here call $\nu\epsilon^{-2/3} \propto
\mathrm{Re_B^{-1}}$. Let us emphasize that $\mathrm{Re_B}$ is distinct from the
hydrodynamic Reynolds number $\mathrm{Re_{NS}}$ for weak advection by a
turbulent Navier-Stokes flow. In such a flow, we have $\alpha_\parallel \sim
\alpha_\perp \sim L$ and $\mathrm{Re_{NS}} \sim \epsilon L
\nu_{\mathrm{NS}}^{-1}$ (where we have now returned to units in which $\langle
v^{-1}\rangle = 1$, so that $\di L = \di T$ and $\epsilon$ is dimensionless).
Assuming near-equality between the flame parameter $\nu$ (arising from thermal
diffusivity) and the hydrodynamic viscosity $\nu_{\mathrm{NS}}$, as is valid
for gaseous combustion, we have $\mathrm{Re_B} \sim \epsilon^{2/3} L
\nu_{\mathrm{NS}}^{-1} \gg \mathrm{Re_{NS}}$ for $\epsilon \ll 1$. Thus the
Burgers turbulence is more fully developed than the hydrodynamic turbulence,
and the Huygens limit $\mathrm{Re_B} \to \infty$ is not incompatible with the
physics of gaseous flames.

The quantitative implications of the white-noise reduction follow from Eq.\
(\ref{dim}). We consider altering a given physical field $\sigma$ (rms value
$\epsilon \ll 1$) by multiplying the correlation length $\alpha_\parallel$ by
$\epsilon^{2/3}$ (compressing in the longitudinal direction) and multiplying
$\sigma$ everywhere by $\epsilon^{-4/3}$. The new field then has an rms value
$\bar\epsilon = \epsilon^{-1/3}$. The parameter $q$ for the new field is
unchanged from the original, and so $\Delta$ is multiplied by
$\epsilon^{-4/3}$, based on the coefficient of $Q(q)$. Thus we have removed the
$\epsilon^{4/3}$ dependence, and the new $\Delta$ is the prefactor. In the
$\epsilon \to 0$ limit, the spectrum of the noise
$\eta(x_\parallel,\mathbf{x}_\perp) = -\epsilon^{-4/3} \sigma(\epsilon^{-2/3}
x_\parallel,\mathbf{x}_\perp)$, obtained by altering the original field
$\sigma$ as described, becomes
\begin{widetext}
\begin{equation}
\label{Deta}
\begin{split}
D(k_\parallel,\mathbf{k}_\perp) &= \int dx_\parallel\, d^{d-1} \mathbf{x}_\perp
\exp(-ik_\parallel x_\parallel - i\mathbf{k}_\perp \cdot \mathbf{x}_\perp)\,
\epsilon^{-8/3} \langle\sigma(0,\mathbf{0})\, \sigma(\epsilon^{-2/3}
x_\parallel,\mathbf{x}_\perp)\rangle\\
&= \int dx_\parallel\, d^{d-1} \mathbf{x}_\perp \exp(-i\mathbf{k}_\perp \cdot
\mathbf{x}_\perp)\, \epsilon^{-2} \langle\sigma(0,\mathbf{0})\,
\sigma(x_\parallel,\mathbf{x}_\perp)\rangle.
\end{split}
\end{equation}
\end{widetext}
Here we have observed that the $-ik_\parallel x_\parallel$ term is negligible
because the correlation function with the argument $\epsilon^{-2/3}
x_\parallel$ vanishes as $\epsilon \to 0$ for all nonzero $x_\parallel$, and we
have then rescaled the integration variable $x_\parallel$. The spectrum is
white in the longitudinal direction because there is no dependence on
$k_\parallel$. The $\epsilon^{-2}$ factor simply normalizes the original
$\sigma$ correlation function to unit variance.

Thus the desired prefactor of $\epsilon^{4/3}$ in the Huygens-front speedup is
the value of $\Delta$ obtained using the path integral (\ref{Feyn}) in the $\nu
\to 0$ limit, with the spectrum (\ref{Deta}) for $\eta$. In other words, the
prefactor is the ground-state binding energy per unit length for the ``directed
polymer'' with energy
\begin{equation}
\label{H}
H = \int du\, [\tfrac{1}{2} |\mathbf{y}'(u)|^2 -
\eta\bm{(}u,\mathbf{y}(u)\bm{)}].
\end{equation}
(The ground-state energy would be zero in the absence of fluctuations, and so
the binding energy is simply the negative of the free energy.) Although this is
the system most commonly referred to as a directed polymer in a random
potential, we observe that the white noise $\eta$ is in no sense small, and
thus $\tfrac{1}{2} |\mathbf{y}'(u)|^2$ is not an accurate approximation to an
incremental arc length. The energy (\ref{H}) is to be considered on its own
terms, separate from the original intuition about a flexible string with
tension. In particular, to use a supposedly more accurate square-root
expression for arc length in Eq.\ (\ref{H}) would be incorrect for our
purposes. Remarkably, the idealizations conventionally adopted to render the
physical directed polymer more tractable (quadratic expansion of arc length,
white-noise potential) are precisely justified in the problem of weakly
perturbed Huygens propagation.

\section{\label{Monotonicity}Monotonicity properties}

\subsection{\label{Spectrum}Dependence on perturbation spectrum}

It is intuitively reasonable that a more vigorous forcing of the Burgers
equation should result in a higher steady-state energy density, and
correspondingly that a random medium with greater fluctuations should result in
faster propagation. More precisely, let us consider the $\nu \to 0$ limit of
the path integral (\ref{partf}) for two different homogeneous weakly random
media, labeled 1 and 2, and assume that at every wave vector the spectrum of
$\sigma_1$ is no greater than that of $\sigma_2$. Because the spectra of
independent random fields are additive, we can obtain a field $\Sigma$ with the
same spectrum as $\sigma_2$ by defining
\begin{equation}
\Sigma = \sigma_1 + \rho,
\end{equation}
where $\rho$ is a zero-mean Gaussian random field (independent of $\sigma_1$)
with an everywhere nonnegative spectrum given by subtraction. According to Eq.\
(\ref{Deta}), as $\epsilon \to 0$, the speedup is governed entirely by the
spectrum (or equivalently the two-point correlation function) of the medium
fluctuations, and so $\Sigma$ gives the same speedup as $\sigma_2$.

Now we observe that the fastest path from the initial front to a given point in
the field $\sigma_1$, which we denote $\mathbf{y}_1(u)$ (with travel time
$t_1$), can be considered as a possible path in the field $\Sigma$. There, its
travel time is different by an amount
\begin{equation}
I = \int_0^{x_\parallel} du\, \rho\bm{(}u,\mathbf{y}_1(u)\bm{)}.
\end{equation}
The fastest path in the field $\Sigma$ must have a travel time no greater than
$t_1 + I$. The ensemble average over $\Sigma$ can then be performed in two
steps, thanks to the statistical independence of $\rho$ and $\sigma_1$. First,
upon averaging over $\rho$ for a given realization of $\sigma_1$, the mean of
$I$ is zero and so the ``partial average'' first-arrival time for $\Sigma$ is
no greater than $t_1$. Second, upon averaging over $\sigma_1$, the mean of
$t_1$ is (obviously) the average first-arrival time for $\sigma_1$.
Consequently, the ``complete average'' first-arrival time for $\Sigma$ (or
$\sigma_2$) is no greater, and the speedup no smaller, than for $\sigma_1$.

This monotonicity property does not immediately relate the prefactors $\Delta$
of $\epsilon^{4/3}$ for different media, because two different spectra with the
same total power $\epsilon^2$ cannot obey the assumed inequality. But relations
can be obtained by considering media with $\epsilon$ differing as little as
possible such that one spectrum is bounded by the other, and then correcting
the result with the known $\epsilon^{4/3}$ scaling. The necessary discrepancy
in $\epsilon$ (which weakens the prefactor relation) can be reduced by
performing an optimally chosen spatial rescaling of one of the media, which
does not affect its Huygens-front speedup.

\begin{table*}
\caption{\label{ss}Correlation functions and normalized spectra of example
two-dimensional random media (all homogeneous and isotropic).}
\begin{ruledtabular}
\begin{tabular}{lll}
Medium &$\langle\sigma(\mathbf{0})\, \sigma(\mathbf{r})\rangle$ &$D(k)$\\\hline
Gaussian (G) &$\epsilon^2 \exp(-r^2/a^2)$ &$\pi a^2 \exp(-\tfrac{1}{4} a^2
k^2)$\\
Modified Gaussian (MG) &$\epsilon^2 (1 - r^2/a^2) \exp(-r^2/a^2)$
&$\tfrac{1}{4} \pi a^4 k^2 \exp(-\tfrac{1}{4} a^2 k^2)$\\
Exponential (E) &$\epsilon^2 \exp(-r/a)$ &$2\pi a^2 (1 + a^2 k^2)^{-3/2}$\\
Modified exponential (ME) &$\epsilon^2 (1 - \tfrac{1}{6} r^2/a^2) \exp(-r/a)$
&$\pi a^4 k^2 (7 + 2 a^2 k^2) (1 + a^2 k^2)^{-7/2}$\\
\end{tabular}
\end{ruledtabular}
\end{table*}

\begin{table}
\caption{\label{comp}Upper bounds (UB) on relative speedup from spectral
monotonicity.}
\begin{ruledtabular}
\begin{tabular}{lllll}
Medium 1 &Medium 2 &$\epsilon_1/\epsilon_2$ &$a_1/a_2$ &UB on
$\Delta_1/\Delta_2$\\\hline
G  &E  &0.723 &1.414 &1.541\\
MG &E  &0.596 &2.000 &1.993\\
MG &ME &0.659 &1.944 &1.743\\
ME &E  &0.904 &1.029 &1.144\\
\end{tabular}
\end{ruledtabular}
\end{table}

Table \ref{ss} lists the correlation functions and spectra of four homogeneous
isotropic random media in $d = 2$ dimensions, which we will use as examples
throughout this paper. Each medium is parametrized by the rms fluctuation
$\epsilon$ and a length scale $a$. We note that the white-noise spectrum
(\ref{Deta}) is just the $d$-dimensional Fourier transform of the normalized
correlation function at the wave vector $\mathbf{k} = (0,\mathbf{k}_\perp)$.
For isotropic media, this is a function $D(|\mathbf{k}_\perp|)$ of exactly the
same form as the Fourier transform $D(k)$ at a general wave vector. The
unnormalized spectrum to be used with the monotonicity property is
$\epsilon^2\, D(k)$. Table \ref{comp} gives the results of numerically
optimizing the relations between pairs of spectra to constrain the speedup
prefactors $\Delta$. The omitted pairs are not useful because one spectrum
dominates as $k \to 0$ and the other as $k \to \infty$.

\subsection{\label{Dimension}Dependence on spatial dimension}

A second monotonicity property applies when a given medium can be viewed as a
``slice'' through a higher-dimensional medium $M$, i.e., as the restriction of
the field $\sigma$ to a (hyper)plane perpendicular to the initial front in $M$.
All paths in the slice are also paths in $M$ with the same travel time, and the
longitudinal coordinate $x_\parallel$ is the same in both media. It follows
that the first-arrival time in $M$ is no greater, and the speedup no smaller,
than in the slice.

For homogeneous isotropic random media, the two-point correlation function of
$\sigma$ in the slice is exactly the same function of distance as in $M$.
Because of the difference in dimensionality, however, its Fourier transform
(i.e., the spectrum) is generally different. Of course, a correlation function
is realizable only if the associated spectrum is nonnegative. We see that if a
given function of distance is a realizable correlation function in $M$, it is
automatically realizable in lower dimensions, and thus the spectrum (although
different) remains nonnegative. But in higher dimensions, the spectrum may
develop negative values and the correlation function may no longer be
realizable.

The most powerful application of this monotonicity property is to a correlation
function that is realizable in all dimensions, such as the Gaussian, whose
Fourier transform is always another Gaussian function. In this case, the
speedup prefactor must be a nondecreasing function of $d$ for all $d \ge 2$.

\subsection{\label{Laminar}Dependence on laminar flame speed}

The special case of Huygens propagation with advection by a random velocity
field, but with a fixed front-advancement speed in the local comoving frame, is
widely studied as a model of turbulent premixed combustion \cite{W85,MK07}.
There, the fixed local speed is called the laminar flame speed $u_L$, and the
overall statistically steady propagation rate is called the turbulent burning
velocity $u_T$. An important question in combustion modeling is the dependence
of $u_T$ on $u_L$ and on the statistics of the advecting flow field (such as
rms velocity $u'$), assumed for simplicity to be given in advance rather than
dynamically affected by the flame. We have shown \cite{MK07} that for $u'/u_L
\ll 1$ (weak turbulence), this idealized combustion problem is equivalent to
weakly random Huygens propagation in a particular quenched medium determined by
the flow statistics (specifically, the two-point spatial correlation function).
Thus the methods of this paper, phrased for convenience in terms of quenched
media, apply also to flames in weak turbulence.

Moreover, a simple monotonicity property allows limited conclusions even about
the opposite limit $u'/u_L \gg 1$ (strong turbulence, the more important case
in practice). Namely, if all flow statistics are held fixed, $u_T$ must be a
nondecreasing function of $u_L$. This holds realization by realization: If we
consider two fronts (initially coincident) propagating independently in the
same flow, the front with smaller $u_L$ can never get ahead of the other front
anywhere, because at the location and time of any such overtaking, it would
have to be advancing faster relative to the flow than its rival.

The connection between weak and strong turbulence is then as follows: For given
flow statistics, the turbulent burning velocity $u_T$ is no greater for very
small $u_L$ (a strong-turbulence problem) than for very large $u_L$ (a
weak-turbulence problem). Numerically this relation is useless, since for very
small $u_L$ we expect $u_T \sim u'$, with a coefficient that remains unknown.
We can only say that $u_T \le \bar u_L$, where $\bar u_L \gg u'$ is a laminar
flame speed large enough that the corresponding turbulent burning velocity is
essentially $\bar u_L$. But the relation is useful in ruling out the
possibility that $u_T$ diverges to infinity at fixed $u'$ as some other flow
parameter is varied. For this to happen in strong turbulence ($u'/u_L \gg 1$),
it would also have to happen in weak turbulence ($u'/u_L \ll 1$), i.e., the
speedup prefactor $\Delta$ would have to diverge. This can be ruled out in a
given case by a suitable upper bound on $\Delta$.

\section{\label{Replica}Replica analysis}

\subsection{Overview of the replica method}

Originally developed in the study of discrete statistical-mechanical systems
such as spin glasses \cite{EA75,BY86,DT05}, the replica method is a powerful
tool for theoretical investigation of the thermodynamics of disordered systems.
It exploits the interplay of two kinds of randomness: the external stochastic
parameters that determine the energy ``landscape'' in which a system finds
itself, and the finite-temperature thermal fluctuations of the system
configuration within that landscape. Although we are here interested mainly in
ground-state properties of the stochastic landscape, it is beneficial to
consider a finite temperature and later take it to zero.

To describe the replica method in the context of our problem
\cite{MP91,MP92,BMP95}, let us denote by $Z(\eta)$ the partition function for
the directed polymer described by Eq.\ (\ref{H}) with a particular realization
of the white-noise potential $\eta(x_\parallel,\mathbf{x}_\perp)$. We are
working with a polymer of a given longitudinal extent at a finite temperature
$\tau$. The most general description of the disordered thermodynamics would
consist of the probability density function (pdf) of $Z(\eta)$ resulting from
the given statistics of $\eta$. This pdf would be difficult to obtain directly,
but its moments can be simplified considerably.

For any positive integer $n$, we have
\begin{equation}
\label{Zn}
Z(\eta)^n = \int \mathcal{D}\{\mathbf{y}_c(u)\}\, \exp\biggl(-\frac{1}{\tau}
\sum_{a=1}^n H\bm{(}\mathbf{y}_a(u)\bm{)}\biggr),
\end{equation}
a product of independent path integrals over $n$ ``replicas'' of the polymer;
here $\mathcal{D}\{\mathbf{y}_c(u)\} \equiv \prod_{c=1}^n
\mathcal{D}\mathbf{y}_c(u)$. Because $H$ is linear in $\eta$, and because
$\eta$ has Gaussian statistics, we can average Eq.\ (\ref{Zn}) using the
identity
\begin{equation}
\label{expzeta}
\langle\exp\zeta\rangle = \exp(\tfrac{1}{2} \langle\zeta^2\rangle),
\end{equation}
valid for any zero-mean Gaussian variable $\zeta$. We identify the Gaussian
variable appearing in the exponential in Eq.\ (\ref{Zn}),
\begin{equation}
\zeta = \frac{1}{\tau} \sum_{a=1}^n \int du\,
\eta\bm{(}u,\mathbf{y}_a(u)\bm{)},
\end{equation}
and compute
\begin{equation}
\begin{split}
\langle\zeta^2\rangle &= \frac{1}{\tau^2} \sum_{a,b=1}^n \int du\, du'\,
\langle\eta\bm{(}u,\mathbf{y}_a(u)\bm{)}\,
\eta\bm{(}u',\mathbf{y}_b(u')\bm{)}\rangle\\
&= \frac{1}{\tau^2} \sum_{a,b=1}^n \int du\, V\bm{(}|\mathbf{y}_a(u) -
\mathbf{y}_b(u)|\bm{)}.
\end{split}
\end{equation}
We have used the relation $\langle\eta(u,\mathbf{y})\,
\eta(u',\mathbf{y}')\rangle = \delta(u - u')\, V(|\mathbf{y} - \mathbf{y}'|)$,
where
\begin{equation}
\label{Vy}
V(y) \equiv \int \frac{d^{d-1} \mathbf{k}_\perp}{(2\pi)^{d-1}}\,
\exp(i\mathbf{k}_\perp \cdot \mathbf{y})\, D(|\mathbf{k}_\perp|).
\end{equation}

Upon combining Eq.\ (\ref{expzeta}) with the part of $H$ that is independent of
$\eta$, we obtain
\begin{equation}
\langle Z(\eta)^n\rangle = \int \mathcal{D}\{\mathbf{y}_c(u)\}\,
\exp\biggl(-\frac{1}{\tau}\, H_n\bm{(}\{\mathbf{y}_c(u)\}\bm{)}\biggr),
\end{equation}
which has the form of a single partition function for $n$ replicas with
``energy''
\begin{equation}
\begin{split}
H_n = \int du\, \biggl(&\tfrac{1}{2} \sum_{a=1}^n |\mathbf{y}'_a(u)|^2\\
&- \frac{1}{2\tau} \sum_{a,b=1}^n V\bm{(}|\mathbf{y}_a(u) -
\mathbf{y}_b(u)|\bm{)}\biggr).
\end{split}
\end{equation}
This $H_n$, unlike $H$, contains no random potential, but the price is that we
now have several interacting polymers (a different number of them for each
different moment of $Z$ we wish to calculate).

Much as in Sec.~\ref{KPZBH}, the new partition function $\langle Z^n\rangle$
can be identified as a Feynman path integral for the $n$-particle
imaginary-time Schr\"odinger equation, with Planck's constant $\tau$ and with
the potential $-V/2\tau$ acting on every pair of particles (including
self-interaction). The potential function is now static and deterministic,
unlike the time-dependent random potential $-\eta$ in the original
Schr\"odinger equation. The wave function
$\psi(\mathbf{y}_1,\dotsc,\mathbf{y}_n)$ is governed by the Hamiltonian
operator
\begin{equation}
\label{Hn}
\mathcal{H}_n = -\tfrac{1}{2} \tau^2 \sum_{a=1}^n \nabla_a^2 - \frac{1}{2\tau}
\sum_{a,b=1}^n V(|\mathbf{y}_a - \mathbf{y}_b|).
\end{equation}
The thermodynamic limit of an infinitely long polymer has a simple
interpretation: By evolving the imaginary-time Schr\"odinger equation
\begin{equation}
\label{itse}
\frac{\partial\psi}{\partial x_\parallel} = -\frac{1}{\tau}\, \mathcal{H}_n
\psi
\end{equation}
for an infinite time, we project any initial wave function onto the quantum
ground state. The path integral for $\langle Z^n\rangle$ over a sufficiently
long time $x_\parallel$ is dominated by a term proportional to $\exp[-E_g(n)\,
x_\parallel/\tau]$, where $E_g(n)$ is the ground-state energy of the
$n$-particle system, because the contributions of the other (higher) energy
eigenvalues are asymptotically negligible.

What we actually wish to calculate is not a moment $\langle Z^n\rangle$ but the
averaged free energy $-\tau\langle\ln Z\rangle$. Unfortunately, $\ln Z$ cannot
be expanded in a Taylor series about $Z = 0$, and to make progress we must
introduce a peculiar feature of the replica method. The needed average is
expressed by the identity
\begin{equation}
\langle\ln Z\rangle = \lim_{n \to 0} \frac{\langle Z^n\rangle - 1}{n},
\end{equation}
and it is assumed that $\langle Z^n\rangle$, computed as above for positive
integer $n$, can be analytically continued to $n$ near zero. Then the
asymptotic (large $x_\parallel$) behavior of the averaged free energy is
\begin{equation}
\begin{split}
{-}\tau\langle\ln Z\rangle &= -\tau\lim_{n \to 0} \frac{\exp[-E_g(n)\,
x_\parallel/\tau] - 1}{n}\\
&= x_\parallel \lim_{n \to 0} \frac{E_g(n)}{n},
\end{split}
\end{equation}
and so the polymer's binding energy per unit length is
\begin{equation}
\Delta = -\lim_{n \to 0} \frac{E_g(n)}{n}.
\end{equation}

\subsection{Variational treatment}

We desire a method for estimating $E_g(n)$ that is demonstrably valid for any
positive integer $n$ and that also (unlike, e.g., numerical solution of the
$n$-particle Schr\"odinger equation) can be formally generalized to noninteger
$n$. A very useful choice is the variational method, which is based on the
observation that the expectation value $\langle\psi|\mathcal{H}_n|\psi\rangle$
of the $n$-particle Hamiltonian in an arbitrary quantum state $|\psi\rangle$ is
an upper bound on $E_g(n)$. To obtain as tight a bound as possible, this
expectation value is minimized over a convenient family of ``trial'' wave
functions, in the hope that some of them are close to the true ground state.

The variational method is valid for positive integer $n$, where $\mathcal{H}_n$
is a Hermitian operator on a well-defined Hilbert space. The wave function
associates a number with each configuration of $n$ particles in $d - 1$
dimensions, but because the Hamiltonian is translation invariant, the center of
mass separates and the nontrivial part of the wave function depends on $n - 1$
vectors. Furthermore, the wave function is subject to one normalization
constraint, $\langle\psi|\psi\rangle = 1$. Thus the number of degrees of
freedom in the wave function can be written
\begin{equation}
f = \infty^{(d-1)(n-1)} - 1.
\end{equation}
When $n = 1$, for example, we have $f = 0$ (no degrees of freedom), as expected
because there is a unique translation-invariant one-particle wave function, the
eigenstate of zero momentum. This variational estimate is automatically the
exact ground state of the free-particle Hamiltonian $\mathcal{H}_1$.

For $n \to 0$, we have $f = -1$, and the replica method relies on the following
nonrigorous argument: If $E_g(n)$ is the minimum of
$\langle\psi|\mathcal{H}_n|\psi\rangle$ with respect to a separate variation in
each degree of freedom of $|\psi\rangle$ around the true ground state, then
when the degrees of freedom are themselves negative in number, any conceivable
variation of $|\psi\rangle$ will result in a \emph{decrease} of
$\langle\psi|\mathcal{H}_n|\psi\rangle$. Consequently, we \emph{maximize}
$\langle\psi|\mathcal{H}_n|\psi\rangle$ among trial wave functions and thereby
obtain a \emph{lower} bound on $E_g(n)$. The results of applying this strategy
to spin glasses have been verified by rigorous methods \cite{G03,T06}, and
there is no evidence that the corresponding results for directed polymers are
invalid. In fact, the numerical simulations in Sec.~\ref{Numerical} provide a
successful quantitative test of this application of the replica method.

In Appendix \ref{appRSB} we analyze a commonly used family of trial wave
functions, parametrized by a function $z(u)$, for which
$\langle\psi|\mathcal{H}_n|\psi\rangle$ can be expressed analytically in $n$.
The result is
\begin{equation}
\begin{split}
\langle\psi|\mathcal{H}_n|\psi\rangle = {}&\tfrac{1}{8} \tau^2 (d - 1)n
\int_1^n \frac{du}{u^2\, \Lambda(u)}\\
&- \frac{n}{2\tau} \left(B_0(0) + \int_1^n du\, B_0\bm{(}z(u)\bm{)}\right),
\end{split}
\end{equation}
where
\begin{align}
\Lambda(u) &= \tfrac{1}{2} u\, z(u) - \tfrac{1}{2} \int_1^u dv\, z(v),\\
B_0(z) &= \int \frac{d^{d-1} \mathbf{k}_\perp}{(2\pi)^{d-1}}\,
\exp(-\tfrac{1}{2} |\mathbf{k}_\perp|^2 z)\, D(|\mathbf{k}_\perp|).
\end{align}
As $n \to 0$, a meaningful wave function requires $z(u)$ to be nonnegative and
nonincreasing for $0 < u < 1$. If we define
\begin{equation}
\label{Gtz}
\begin{split}
\Gamma(\tau,z) = {}&-\lim_{n \to 0}
\frac{\langle\psi|\mathcal{H}_n|\psi\rangle}{n}\\
= {}&\tfrac{1}{8} \tau^2 (d - 1) \int_0^1 \frac{du}{u^2\, \Lambda(u)}\\
&+ \frac{1}{2\tau} \left(B_0(0) - \int_0^1 du\, B_0\bm{(}z(u)\bm{)}\right),
\end{split}
\end{equation}
then the polymer's binding energy per unit length at temperature $\tau$ is
bounded above by $\Gamma(\tau,z)$ for any nonnegative, nonincreasing function
$z(u)$. Thus the Huygens prefactor $\Delta$ obeys
\begin{equation}
\Delta \le \lim_{\tau \to 0} \Gamma(\tau, z_\tau),
\end{equation}
where we anticipate based on previous results \cite{B94} that a useful, finite
bound will require a $\tau$-dependent choice of $z(u)$.

\subsection{\label{Explicit}Explicit replica bounds}

The replica variational treatment has been applied in some detail to the
directed-polymer problem, focusing on the case of the Gaussian medium
\cite{B94}. In that work, all values of temperature were considered, and the
goal was a general qualitative understanding of the polymer's behavior, rather
than a calculation of its binding energy. Here, by concentrating on the binding
energy in the zero-temperature limit, we are able to extend the previous
results to obtain explicit bounds on $\Delta$, not only for the Gaussian medium
but for arbitrary spectra. (``Gaussian'' refers to a type of spectrum, defined
for $d = 2$ in Table \ref{ss}. As discussed in Sec.~\ref{White}, a quite
separate feature of Huygens propagation is the equivalence of weak isotropic
perturbations, not necessarily Gaussian in pdf, to the directed polymer's
white-noise perturbations, which are automatically Gaussian in pdf and
described completely by their spectrum.)

We start with a derivation of the equation for stationarity of $\Gamma(\tau,z)$
and its consequences, along the lines of the previous work \cite{B94}. From
Eq.\ (\ref{Lambdaz}), we compute
\begin{equation}
\frac{\delta\Lambda(v)}{\delta z(u)} = \tfrac{1}{2} u\, \delta(u - v) +
\tfrac{1}{2}\, \theta(u - v),
\end{equation}
where $\theta(x)$ equals $1$ for $x > 0$ and $0$ for $x < 0$, and thus
\begin{widetext}
\begin{equation}
\label{dGdz}
\begin{split}
\frac{\delta\Gamma(\tau,z)}{\delta z(u)} &= -\tfrac{1}{16} \tau^2 (d - 1)
\int_0^1 \frac{dv}{v^2\, \Lambda(v)^2} [u\, \delta(u - v) + \theta(u - v)] -
\frac{1}{2\tau}\, B_0'\bm{(}z(u)\bm{)}\\
&= -\tfrac{1}{16} \tau^2 (d - 1) \left(\frac{1}{u\, \Lambda(u)^2} + \int_0^u
\frac{dv}{v^2\, \Lambda(v)^2}\right) + \frac{1}{2\tau}\, B_1\bm{(}z(u)\bm{)}.
\end{split}
\end{equation}
\end{widetext}
We use the notation
\begin{equation}
\label{Bp}
\begin{split}
B_p(z) &\equiv (-1)^p\, \frac{d^p B_0(z)}{dz^p}\\
&= \int \frac{d^{d-1}\mathbf{k}_\perp}{(2\pi)^{d-1}}\, (\tfrac{1}{2}
|\mathbf{k}_\perp|^2)^p \exp(-\tfrac{1}{2} |\mathbf{k}_\perp|^2 z)\,
D(|\mathbf{k}_\perp|);
\end{split}
\end{equation}
note that each $B_p(z)$ is a positive, decreasing function for $z > 0$.

In the absence of constraints on $z(u)$, the variational optimum would satisfy
$\delta\Gamma/\delta z = 0$. Inserting Eq.\ (\ref{dGdz}) and differentiating
with respect to $u$, we obtain
\begin{equation}
\tfrac{1}{8} \tau^2 (d - 1)\, \frac{\Lambda'(u)}{u\, \Lambda(u)^3} -
\frac{1}{2\tau}\, B_2\bm{(}z(u)\bm{)}\, z'(u) = 0,
\end{equation}
or, using Eq.\ (\ref{Lpzp}),
\begin{equation}
\label{var}
z'(u) \left(\frac{\tau^2 (d - 1)}{16\, \Lambda(u)^3} -
\frac{B_2\bm{(}z(u)\bm{)}}{2\tau}\right) = 0.
\end{equation}
Regions where $z'(u) = 0$ (for which infinitesimal variations could violate the
nonincreasing constraint) still obey Eq.\ (\ref{var}). Provided $z(u) > 0$
everywhere (so that the nonnegative constraint has no local effect), Eq.\
(\ref{var}) is a necessary condition for an optimum.

The type of solution obtained for the $d = 2$ Gaussian medium \cite{B94} has
the second factor in Eq.\ (\ref{var}) equal to zero for $0 < u < \uc$, while
$z(u) = z(\uc)$ for $\uc < u < 1$ so that the first factor $z'(u)$ equals zero
there. Thus we have
\begin{align}
\label{Luluc}
\Lambda(u) &= \tfrac{1}{2} \tau \left(\frac{d -
1}{B_2\bm{(}z(u)\bm{)}}\right)^{1/3} & &(0 < u < \uc),\\
\Lambda(u) &= \tfrac{1}{2}\, z(\uc) & &(\uc < u < 1).
\end{align}
From the assumed continuity of $z(u)$ at $\uc$, the continuity of $\Lambda(u)$
follows, and so
\begin{equation}
\label{Luc}
\tfrac{1}{2}\, z(\uc) = \Lambda(\uc) = \tfrac{1}{2}\tau \left(\frac{d -
1}{B_2\bm{(}z(\uc)\bm{)}}\right)^{1/3}.
\end{equation}
As $\tau \to 0$, this gives
\begin{equation}
\label{zuc}
z(\uc) = \tau \left(\frac{d - 1}{B_2(0)}\right)^{1/3}.
\end{equation}
Also, by differentiating Eq.\ (\ref{Luluc}), we find
\begin{align}
\Lambda'(u) &= \tfrac{1}{6}\tau\, \frac{(d -
1)^{1/3}}{B_2\bm{(}z(u)\bm{)}^{4/3}}\, B_3\bm{(}z(u)\bm{)}\, z'(u) & &(0 < u <
\uc),
\end{align}
or, assuming $z'(u) \neq 0$ and using Eq.\ (\ref{Lpzp}),
\begin{align}
\label{uzu}
u &= \tfrac{1}{3}\tau\, \frac{(d - 1)^{1/3}}{B_2\bm{(}z(u)\bm{)}^{4/3}}\,
B_3\bm{(}z(u)\bm{)} & &(0 < u < \uc).
\end{align}
This constitutes an implicit solution for $z(u)$. We would like to make use of
this solution as far as possible for arbitrary spectra, without adopting
particular forms of $B_p$.

Let us define
\begin{equation}
\label{muz}
\mu(z) = \frac{B_3(z)}{B_2(z)^{4/3}}.
\end{equation}
The simplest case occurs when $\mu(z)$ is a decreasing function for all $z > 0$
and $\lim_{z \to \infty} \mu(z) = 0$. Then a single-valued, decreasing function
$z(u)$ for $0 < u < \uc$ is defined by
\begin{align}
\label{uz}
u &= \tfrac{1}{3} \tau (d - 1)^{1/3}\, \mu(z) & &[z(\uc) < z < \infty],
\end{align}
where $z(\uc)$ is given by Eq.\ (\ref{zuc}). We now attempt to substitute this
trial solution into Eq.\ (\ref{Gtz}). Upon integration by parts, the first term
of $\Gamma(\tau,z)$ becomes
\begin{equation}
\label{G1}
\begin{split}
\Gamma_1 = {}&\tfrac{1}{8} \tau^2 (d - 1) \left(\frac{1}{\Lambda(\uc)}
\int_{\uc}^1 \frac{du}{u^2} + \int_0^{\uc} \frac{du}{u^2\, \Lambda(u)}\right)\\
= {}&\tfrac{1}{8} \tau^2 (d - 1)\\
&\times \left(\frac{1/\uc - 1}{\Lambda(\uc)} - \int_0^{\uc} du\,
\frac{\Lambda'(u)}{u\, \Lambda(u)^2} - \left.\frac{1}{u\,
\Lambda(u)}\right|_0^{\uc}\right)\\
= {}&\tfrac{1}{8} \tau^2 (d - 1)\\
&\times \left(-\frac{1}{\Lambda(\uc)} - \tfrac{1}{2} \int_0^{\uc} du\,
\frac{z'(u)}{\Lambda(u)^2} + \lim_{u \to 0} \frac{1}{u\, \Lambda(u)}\right).
\end{split}
\end{equation}
To evaluate the limit, we observe that $u \to 0$ corresponds to $z \to \infty$.
If $D(k) \sim k^q$ as $k \to 0$, then for the total power
\begin{equation}
\int \frac{d^d \mathbf{k}}{(2\pi)^d}\, D(k)
\end{equation}
to be finite, as required to define the parameter $\epsilon$ of the random
medium, we must have $q + d > 0$. The resulting behavior of $B_p$, from Eq.\
(\ref{Bp}), is
\begin{align}
\label{Bpasym}
B_p(z) &\sim z^{-p - (q + d - 1)/2} & &(z \to \infty).
\end{align}
Using Eqs.\ (\ref{Luluc}) and (\ref{uz}), we find
\begin{align}
u\, \Lambda(u) &\propto \frac{B_3(z)}{B_2(z)^{5/3}} \sim z^{(q + d)/3} \to
\infty & &(z \to \infty),
\end{align}
and so the limit in Eq.\ (\ref{G1}) is zero. Furthermore, from Eq.\
(\ref{Luc}), the term $\propto \tau^2/\Lambda(\uc)$ in Eq.\ (\ref{G1}) scales
with $\tau$ and vanishes as $\tau \to 0$. Under a change of variable to $z$,
the remaining integral gives
\begin{equation}
\label{G1a}
\Gamma_1 = \tfrac{1}{4} (d - 1)^{1/3} \int_{z(\uc)}^\infty dz\, B_2(z)^{2/3}.
\end{equation}

The second term of $\Gamma(\tau,z)$ becomes
\begin{equation}
\label{G2}
\begin{split}
\Gamma_2 = \frac{1}{2\tau}\, \biggl(&B_0(0) - B_0\bm{(}z(\uc)\bm{)}
\int_{\uc}^1 du\\
&- \int_0^{\uc} du\, B_0\bm{(}z(u)\bm{)}\biggr)\\
= \frac{1}{2\tau}\, \biggl(&B_0(0) - (1 - \uc)\, B_0\bm{(}z(\uc)\bm{)}\\
&+ \int_{z(\uc)}^\infty dz\, B_0(z)\, u'(z)\biggr),
\end{split}
\end{equation}
where $u(z)$ is given by Eq.\ (\ref{uz}). Integration by parts then yields
\begin{equation}
\Gamma_2 = \frac{1}{2\tau}\, \biggl(B_0(0) - B_0\bm{(}z(\uc)\bm{)} +
\int_{z(\uc)}^\infty dz\, B_1(z)\, u(z)\biggr),
\end{equation}
because $\lim_{z \to \infty} B_0(z)\, u(z) = 0$. Recognizing a difference
quotient, which as $\tau \to 0$ becomes a derivative $\propto B_0'(0)$, we
obtain
\begin{equation}
\label{G2a}
\begin{split}
\Gamma_2 = {}&\tfrac{1}{2}\, B_1(0) \left(\frac{d - 1}{B_2(0)}\right)^{1/3}\\
&+ \tfrac{1}{6} (d - 1)^{1/3} \int_{z(\uc)}^\infty dz\, \frac{B_1(z)\,
B_3(z)}{B_2(z)^{4/3}}.
\end{split}
\end{equation}
In both Eqs.\ (\ref{G1a}) and (\ref{G2a}), the lower limits of the integrals
can be taken to zero by Eq.\ (\ref{zuc}), since there is no remaining singular
dependence on $\tau$. Also, because
\begin{equation}
\frac{d}{dz}\, \frac{B_1(z)}{B_2(z)^{1/3}} = -B_2(z)^{2/3} + \tfrac{1}{3}\,
\frac{B_1(z)\, B_3(z)}{B_2(z)^{4/3}},
\end{equation}
Eq.\ (\ref{G2a}) simplifies upon a further integration by parts to
\begin{equation}
\begin{split}
\Gamma_2 = {}&\tfrac{1}{2} (d - 1)^{1/3} \int_0^\infty dz\, B_2(z)^{2/3}\\
&+ \tfrac{1}{2} (d - 1)^{1/3} \lim_{z \to \infty} \frac{B_1(z)}{B_2(z)^{1/3}}.
\end{split}
\end{equation}
The limit vanishes because
\begin{align}
\frac{B_1(z)}{B_2(z)^{1/3}} &\sim z^{-(q + d)/3} \to 0 & &(z \to \infty).
\end{align}
Thus we obtain the replica bound
\begin{equation}
\label{RB}
\Delta \le \Gamma_1 + \Gamma_2 = \tfrac{3}{4} (d - 1)^{1/3} \int_0^\infty dz\,
B_2(z)^{2/3}
\end{equation}
on the prefactor of the Huygens-front speedup.

A remarkable renormalization interpretation of Eq.\ (\ref{RB}) is seen by
rewriting it as
\begin{equation}
\label{RBa}
\Delta \le \tfrac{3}{4} (d - 1)^{1/3} \int_0^\infty \frac{dz}{z}\, [z^{3/2}\,
B_2(z)]^{2/3}.
\end{equation}
Note that
\begin{equation}
\label{z32B}
\begin{split}
z^{3/2}\, B_2&(z)\\
= {}&\int \frac{d^{d-1} \mathbf{k}_\perp}{(2\pi)^{d-1}}\, \tfrac{1}{4}
|\mathbf{k}_\perp|^4\, z^{3/2} \exp(-\tfrac{1}{2} |\mathbf{k}_\perp|^2 z)\,
D(|\mathbf{k}_\perp|)\\
= {}&\tfrac{1}{2} \pi^{1/2}\, \frac{\Gamma(\frac{1}{2}
d)}{\Gamma\bm{(}\frac{1}{2} (d - 1)\bm{)}}\\
&\times \int \frac{d^d \mathbf{k}}{(2\pi)^d}\, (k^2 z)^{3/2} \exp(-\tfrac{1}{2}
k^2 z)\, D(k),
\end{split}
\end{equation}
where we have inserted factors to compensate increasing the integration from $d
- 1$ to $d$ dimensions. Because the integrand becomes small for $k \ll
z^{-1/2}$ or $k \gg z^{-1/2}$, Eq.\ (\ref{z32B}) represents the power contained
in a finite wave-number band around $k \sim z^{-1/2}$ of width $\delta k \sim
z^{-1/2}$. In Eq.\ (\ref{RBa}), this spectral band power (analogous to
$\epsilon^2$) is raised to the $\frac{2}{3}$ power (consistent with
$\epsilon^{4/3}$ scaling) and then integrated over all logarithmic length
scales ($dz/z$). This suggests a stepwise process in which, starting from the
smallest length scales, each order-unity spectral band has the same qualitative
effect as if acting alone: It renormalizes the front propagation speed, with
the new effective speed (turbulent burning velocity in combustion) providing
the raw input (laminar flame speed) for the next larger-length-scale band. In
the weak-perturbation limit of Huygens propagation, because all these
renormalization contributions are very small, they combine additively as
displayed in Eq.\ (\ref{RBa}). These conclusions are compared to existing
concepts of front-speed renormalization in Sec.~\ref{Discussion}.

The assumptions about $\mu(z)$ stated below Eq.\ (\ref{muz}) hold for the
two-dimensional Gaussian and exponential media in Table \ref{ss}. As a result,
bounds on their speedup prefactors can be obtained from Eq.\ (\ref{RB}):
\begin{align}
\label{DG}
\Delta_{\text{G}} &\le \frac{3^{8/3} \pi^{1/3}}{16} \simeq 1.714,\\
\label{DE}
\Delta_{\text{E}} &\le 2.038.
\end{align}
For the exponential medium, due to the long spectral tail at high wave number,
$B_p(0)$ is divergent for $p \ge 1$. Although such quantities appeared in the
derivation, the exponential medium can be approached by a limiting process to
make the expressions well-defined. The end result, Eq.\ (\ref{RB}), is finite
and was evaluated directly by numerical integration to obtain Eq.\ (\ref{DE}).

We now consider violations of the previous assumptions about $\mu(z)$. If
$\mu(z)$ is decreasing for all $z > 0$ but $\lim_{z \to \infty} \mu(z) = \mu_*
> 0$, then we define $z(u)$ by Eq.\ (\ref{uz}) for $\frac{1}{3} \tau (d -
1)^{1/3} \mu_* \equiv u_* < u < \uc$, and define $z(u) = \infty$ for $0 < u <
u_*$. It follows that $\Lambda(u) = \infty$ for $0 < u < u_*$, and so both
integrals in Eq.\ (\ref{Gtz}) receive nonzero contributions only from $u_* < u
< 1$, since $B_0(\infty) = 0$. The change of variable to $z$ and the
integrations by parts proceed as before, and the result (\ref{RB}) is
unchanged.

On the other hand, if $\mu(z)$ is not an everywhere decreasing function, let
$[0,z_*]$ be the largest interval from zero on which it is decreasing (possibly
$z_* = 0$). We define $u_* = \frac{1}{3} \tau (d - 1)^{1/3}\, \mu(z_*)$ and
again take $z(u) = \Lambda(u) = \infty$ for $0 < u < u_*$. Then the $z$
integrals extend only up to $z_*$, and several boundary terms from integration
by parts no longer vanish. Specifically, we find
\begin{align}
\begin{split}
\Gamma_1 = {}&\tfrac{1}{8} \tau^2 (d - 1)\, \frac{1}{u_*\, \Lambda(u_*)}\\
&+ \tfrac{1}{4} (d - 1)^{1/3} \int_0^{z_*} dz\, B_2(z)^{2/3},
\end{split}\\
\begin{split}
\Gamma_2 = {}&\frac{1}{2\tau}\, B_0(z_*)\, u_* + \tfrac{1}{2} (d - 1)^{1/3}\,
\frac{B_1(z_*)}{B_2(z_*)^{1/3}}\\
&+ \tfrac{1}{2} (d - 1)^{1/3} \int_0^{z_*} dz\, B_2(z)^{2/3},
\end{split}
\end{align}
and thus
\begin{equation}
\label{RBb}
\begin{split}
\Delta \le {}&\Gamma_1 + \Gamma_2\\
= {}&(d - 1)^{1/3}\, \biggl(\tfrac{3}{4}\, \frac{B_2(z_*)^{5/3}}{B_3(z_*)} +
\tfrac{1}{6}\, \frac{B_0(z_*)\, B_3(z_*)}{B_2(z_*)^{4/3}}\\
&+ \tfrac{1}{2}\, \frac{B_1(z_*)}{B_2(z_*)^{1/3}} + \tfrac{3}{4} \int_0^{z_*}
dz\, B_2(z)^{2/3}\biggr).
\end{split}
\end{equation}
This bound applies to the two remaining media in Table \ref{ss}. For the
modified Gaussian medium, $\mu(z)$ is in fact an \emph{increasing} function for
$z > 0$, so we take $z_* = 0$, giving
\begin{equation}
\label{DMG}
\Delta_{\text{MG}} \le \frac{359\pi^{1/3}}{2^{8/3} 3^{4/3} 5^{1/3} 7} \simeq
1.599.
\end{equation}
For the modified exponential medium, numerical evaluation shows that $\mu(z)$
is decreasing only for $0 < z < z_* \simeq 7.492$, giving
\begin{equation}
\label{DME}
\Delta_{\text{ME}} \le 1.943.
\end{equation}
The reason $\mu(z)$ ultimately increases for these media is that $D(k) \sim
k^2$ as $k \to 0$. From Eq.\ (\ref{Bpasym}) with $q = 2$ and $d = 2$, we find
that $\mu(z) \sim z^{1/6}$ as $z \to \infty$. By contrast, two-dimensional
media with $D(k) \sim k^0$ as $k \to 0$ have $\mu(z) \sim z^{-1/6}$ as $z \to
\infty$.

It is possible to obtain a slightly better bound on $\Delta_{\text{MG}}$.
Because $z_* = 0$ for this medium, no real use has been made of the solution
(\ref{uzu}). The trial functions we have constructed are simply piecewise
constant, of the form
\begin{align}
\label{ul1s}
z(u) &= \infty, & \Lambda(u) &= \infty & &(0 < u < \uc),\\
\label{ug1s}
z(u) &= \zc, & \Lambda(u) &= \tfrac{1}{2} \zc & &(\uc < u < 1).
\end{align}
We can discard the earlier motivation and consider arbitrary permissible values
of $\uc$ and $\zc$. Upon variational optimization, Eqs.\ (\ref{ul1s}) and
(\ref{ug1s}) are known as the one-step solution because only one level of
replica symmetry breaking is needed ($K = 1$, $m_1 = \uc$). This solution was
previously discussed for the directed polymer with arbitrary perturbation
spectrum \cite{G93} and will now be derived in our notation. Equation
(\ref{Gtz}) becomes
\begin{equation}
\label{G1s}
\begin{split}
\Gamma(\tau,z) = {}&\tfrac{1}{8} \tau^2 (d - 1) \left(\frac{1}{\uc} - 1\right)
\frac{2}{\zc}\\
&+ \frac{1}{2\tau}\, [B_0(0) - (1 - \uc)\, B_0(\zc)].
\end{split}
\end{equation}
Stationarity with respect to $\uc$ and $\zc$ gives
\begin{align}
-\tfrac{1}{4} \tau^2 (d - 1)\, \frac{1}{\uc^2 \zc} + \frac{B_0(\zc)}{2\tau} &=
0,\\
-\tfrac{1}{4} \tau^2 (d - 1)\, \left(\frac{1}{\uc} - 1\right) \frac{1}{\zc} +
(1 - \uc)\, \frac{B_1(\zc)}{2\tau} &= 0.
\end{align}
As $\tau \to 0$, the solution is
\begin{align}
\uc &= \tau \left(\frac{d - 1}{2}\right)^{1/3}\,
\frac{B_1(0)^{1/3}}{B_0(0)^{2/3}},\\
\zc &= \tau \left(\frac{d - 1}{2}\right)^{1/3}\,
\frac{B_0(0)^{1/3}}{B_1(0)^{2/3}},
\end{align}
and substituting into Eq.\ (\ref{G1s}) shows that
\begin{equation}
\label{RB1}
\Delta \le \tfrac{3}{2} \left(\frac{d - 1}{2}\right)^{1/3}\, B_0(0)^{1/3}\,
B_1(0)^{1/3}.
\end{equation}

For the two-dimensional modified Gaussian medium, Eq.\ (\ref{RB1}) gives a
tighter bound than Eq.\ (\ref{DMG}),
\begin{equation}
\label{DMG1}
\Delta_{\text{MG}} \le \frac{3^{4/3} \pi^{1/3}}{4} \simeq 1.585,
\end{equation}
as expected because we have optimized over a new family that includes our
previous trial solution. For the Gaussian medium, we obtain a looser bound than
before ($\Delta_{\text{G}} \le 3\pi^{1/3}/2^{4/3} \simeq 1.744$); for the
exponential and modified exponential media, since $B_1(0)$ diverges, the
one-step upper bound is infinite and uninformative.

\subsection{Implications of the replica results}

We now discuss the replica results in light of the monotonicity properties of
Sec.~\ref{Monotonicity}. The finite-band-renormalization interpretation
described below Eq.\ (\ref{RB}) indicates that media with broader spectra (on a
logarithmic wave-number scale) should have larger bounds on $\Delta$. This is
because the spectral power is more widely dispersed among bands, giving a lower
amount per band before each is raised to the $\frac{2}{3}$ power, and $x^{2/3}$
decreases more slowly than $x$ as $x$ becomes small. [The limit $x \to 0$
corresponds to a spectrum $D(k) \propto k^{-d}$ with power spread equally over
many orders of magnitude in wave number.] Indeed, the replica bounds on
$\Delta$ are larger for the ``multiscale'' media E and ME, which have long
spectral tails at high wave number, than for the ``single-scale'' media G and
MG, whose spectra fall off very rapidly at high wave number. Furthermore, each
``modified'' medium, exhibiting a suppression of low wave numbers, has a
smaller replica bound than the medium from which it was derived. Consequently,
the spectral-monotonicity bounds given in Table \ref{comp}, though valid, are
not particularly sharp. Those bounds involve scaling down the amplitude of the
narrower spectrum (medium~1) until it fits under the broader spectrum
(medium~2). The resulting upper bound on $\Delta_1/\Delta_2$ is necessarily
greater than unity, whereas the true value is expected to be less than unity if
these inferences based on the replica bounds are accurate, a hypothesis that is
tested numerically in Sec.~\ref{Numerical}.

The dependence on spatial dimension discussed in Sec.~\ref{Dimension} is
confirmed by the replica bounds for Gaussian media for various $d$. Although in
Sec.~\ref{Explicit} we numerically computed the replica bounds only for certain
two-dimensional media, we emphasize that Eqs.\ (\ref{RB}), (\ref{RBb}), and
(\ref{RB1}) are valid for all $d \ge 2$ under the stated assumptions and
definitions. Gaussian media are the most straightforward to consider for
arbitrary $d$ because the spectrum and the spatial correlation function can
both retain a Gaussian form. (The other media in Table \ref{ss} are inherently
two-dimensional; extending them to $d > 2$ would be a matter of definition and
would require qualitatively changing either the correlation function or the
spectrum, the former possibly departing from monotonicity, the latter
jeopardizing realizability.) The normalized Gaussian spectrum is the
$d$-dimensional Fourier transform of $\exp(-r^2/a^2)$, i.e.,
\begin{equation}
\begin{split}
D(k) &= \int d^d \mathbf{r} \exp(-i\mathbf{k} \cdot \mathbf{r})
\exp\biggl(-\frac{r^2}{a^2}\biggr)\\
&= \pi^{d/2} a^d \exp(-\tfrac{1}{4} a^2 k^2).
\end{split}
\end{equation}
For $d \ge 3$, we find that $\mu(z)$ is a nondecreasing function (much as for
the modified Gaussian medium in $d = 2$), and we are driven to the one-step
solution. Equation (\ref{RB1}) gives
\begin{equation}
\Delta \le \frac{3\pi^{1/3}}{2^{4/3}}\, (d - 1)^{2/3} \simeq 1.744 (d -
1)^{2/3},
\end{equation}
consistent with $\Delta$ being an increasing function of $d$. [The extra factor
$(d - 1)^{1/3}$ comes from the $d$-dependence of $B_p$.] The one-step replica
solution for Gaussian media with $d \ge 3$ was previously discussed from the
perspective of white-noise-driven Burgers turbulence \cite{BMP95}.

Finally, the replica bounds, in combination with the link between weak and
strong random advection in Sec.~\ref{Laminar}, have an important implication
for idealized turbulent combustion. Weak-turbulence bounds follow from the
relation between weakly random quenched and advected media mentioned in
Sec.~\ref{White}. Specifically, an isotropic incompressible random flow, with
kinetic energy per unit wave number $E(k)$, is equivalent in speedup to an
isotropic quenched medium with unnormalized spectrum
\begin{equation}
\label{EtoD}
\epsilon^2\, D(k) = \frac{2^d \pi^{d/2}\, \Gamma(\frac{1}{2} d)}{d - 1}\,
\frac{E(k)}{k^{d-1}}.
\end{equation}
This relation leads to $\epsilon^2 = u'^2/(d - 1)$ in terms of the mean square
velocity $u'^2 = 2 \int_0^\infty dk\, E(k)$---as expected because at each
relevant wave vector $\mathbf{k}$ (with $k_\parallel = 0$), the velocity
fluctuations are distributed over $d - 1$ directions transverse to
$\mathbf{k}$, only one of which (the $x_\parallel$ direction) contributes to
the speedup.

The Kolmogorov spectrum of Navier-Stokes turbulence in the limit of infinite
$\mathrm{Re}$ has $E(k) \sim k^{d+1}$ for $k \to 0$ and $E(k) \sim k^{-5/3}$
for $k \to \infty$. We can take $\mathrm{Re} = u'L/\nu_{\mathrm{NS}} \to
\infty$ by reducing $\nu_{\mathrm{NS}}$ or---since the speedup is insensitive
to spatial rescaling---by increasing the integral scale $L$ at fixed
$\nu_{\mathrm{NS}}$. This Kolmogorov spectrum is qualitatively similar to that
of the modified exponential medium. The effective $D(k)$ for turbulence is
proportional to $k^2$ for $k \to 0$ (matching medium ME) and to $k^{-2/3 - d}$
for $k \to \infty$ [i.e., a one-dimensional spectrum $k^{d - 1}\, D(k) \sim
k^{-5/3}$, versus $k^{-2}$ for medium ME]. Just as with medium ME, $B_p(0)$
diverges in $\mathrm{Re} = \infty$ turbulence for $p \ge 1$, so the one-step
bound on the flame speedup is uninformative. But Eq.\ (\ref{RBb}) provides a
finite upper bound, because the Kolmogorov spectrum gives $B_2(z) \sim
z^{-7/6}$ as $z \to 0$ (for all $d$) and thus the integral of $B_2(z)^{2/3}$
converges. The value of $z_*$ is finite because $\mu(z)$ decreases for small
$z$ [$\mu(z) = -B_2'/B_2^{4/3} \sim z^{-11/18}$] and then increases for large
$z$ [$\mu(z) \sim z^{(d - 1)/6}$ from Eq.\ (\ref{Bpasym})]. We conclude that
the weak-turbulence speedup is finite even for $\mathrm{Re} = \infty$. Since
the turbulent burning velocity $u_T$ is a nondecreasing function of the laminar
flame speed $u_L$, it follows that $u_T$ remains finite for $\mathrm{Re} =
\infty$ in the case of strong turbulence ($u_L \to 0$ for a given flow, i.e.,
fixed $u'$). The importance of this result for combustion modeling will be
discussed in a future publication.

\section{\label{Numerical}Numerical tests}

\subsection{\label{Existing}Existing simulations}

Here we review the available data that quantitatively describe the behavior of
a relevant system (one of the class of equivalent problems discussed in
Sec.~\ref{Relations}) and can be compared directly with our analytical results.
These data are from numerical simulations with $d = 2$: either two-dimensional
weakly random Huygens propagation or one-dimensional white-noise-driven Burgers
turbulence. Existing experimental front-propagation results are not
sufficiently reliable for comparison due to the difficulty of approaching all
the required idealizations (pure Huygens propagation, weak perturbations,
unbounded statistically homogeneous medium). Even $d = 2$ experiments that
appear to confirm the $\epsilon^{4/3}$ speedup scaling \cite{SMVPCRSS98} do not
correspond to the simple white-noise reduction of Sec.~\ref{White}, because the
medium is artificially constructed from statistically independent patches on a
regular grid aligned with the propagation direction, producing long-range
correlations in the medium structure. Numerical simulations of
three-dimensional Huygens propagation \cite{KA92,KA94} also yield
$\epsilon^{4/3}$ scaling but involve similar long-range correlations in
addition to transverse anisotropy.

Comparisons with two of our example spectra can be made for existing
simulations of Huygens propagation in two-dimensional isotropic quenched media,
motivated by applications in seismology \cite{RMS93}. Medium G and medium ME
(there called simply ``exponential'') are synthesized as Gaussian random
fields, with $\epsilon$ ranging from $0.005$ to $0.1$, and travel times are
computed by an algorithm based on Huygens' principle. Plots show the expected
transient growth of the speedup, with a significant (but not yet complete)
leveling-off at the longest propagation distances. Thus the results obtained
should underestimate the steady-state speedup, and definitely be below the
replica bounds. The power laws reported from fitting the speedup are
$0.0026(100\epsilon)^{1.33}$ (medium G) and $0.0035(100\epsilon)^{1.26}$
(medium ME). Adjusting the results to an $\epsilon^{4/3}$ law based on a
central value $\epsilon = 0.02$ to obtain the best estimate of the prefactor,
we find
\begin{align}
\Delta_{\text{G}} &\ge 1.2,\\
\Delta_{\text{ME}} &\ge 1.5,
\end{align}
where the inequalities reflect the incomplete equilibration. Indeed, these
lower bounds are consistent with, and reasonably close to, the replica upper
bounds (\ref{DG}) and (\ref{DME}).

Useful results for medium MG are available from high-resolution numerical
simulations \cite{GK98,G99} of the one-dimensional viscous Burgers equation
with white-in-time forcing at $\mathrm{Re_B} \sim 10^4$ (very close to the
inviscid limit), where $\mathrm{Re_B}$ is the Burgers-fluid Reynolds number
defined in Sec.~\ref{White}. The spatial forcing spectrum corresponds to medium
MG, but, as with other Burgers simulations focusing on universal features like
small-scale structure functions and velocity pdf tails, the key nonuniversal
parameters (forcing amplitude, energy density) are reported only roughly. Using
raw simulation data \cite{G06}, however, we determine the steady-state energy
density
\begin{equation}
\label{ww}
\tfrac{1}{2} \langle w^2\rangle = 1.60(20),
\end{equation}
where the $1\sigma$ statistical uncertainty is estimated by dividing the data
into three segments, and is substantial because only a few ``large-eddy
turnover times'' are simulated in a steady state. (This is an appropriate
tradeoff for simulations focusing on small-scale features and thus requiring
high resolution.) Equation (\ref{ww}) applies for a forcing spectrum
\cite{G99,G06} with $a = 4 \times 10^4$ and
\begin{equation}
\int_{-\infty}^\infty \frac{dk_\perp}{2\pi}\, \tfrac{1}{2} k_\perp^4\,
D(|k_\perp|) = 5 \times 10^{-13} = \frac{32}{a^3}
\end{equation}
(exact values), whereas the normalized MG spectrum in Table \ref{ss} has this
integral equal to $15\pi^{1/2}/a^3$. Correcting Eq.\ (\ref{ww}) with the factor
$(15\pi^{1/2}/32)^{2/3}$, we obtain
\begin{equation}
\Delta_{\text{MG}} = 1.42(17).
\end{equation}
This result suggests that the replica bound (\ref{DMG1}) is valid, and if
valid, it is seen to be fairly sharp.

\subsection{New simulations}

To obtain high-precision values for the speedup prefactor in our four example
media, we have developed a geometric algorithm for numerical evolution of the
one-dimensional inviscid KPZ equation
\begin{equation}
\label{KPZ1}
\frac{\partial h}{\partial t} = \tfrac{1}{2} \left(\frac{\partial h}{\partial
x}\right)^2 + \eta(t,x)
\end{equation}
with white-in-time forcing. This can be simulated more efficiently than the
original propagation problem, because for very small $\epsilon$, the
white-noise process in $t$ reflects a longitudinal distance scale of front
evolution that is much longer than the correlation length of the medium
\cite{KA94}. To define the problem precisely, we assume standard periodic
boundary conditions on a lateral domain $0 \le x \le L$, with bulk properties
recovered in the limit $L \to \infty$. Details of the numerical method are
given in Appendix \ref{appNumerical}.

\begin{figure}
\includegraphics[width=86mm]{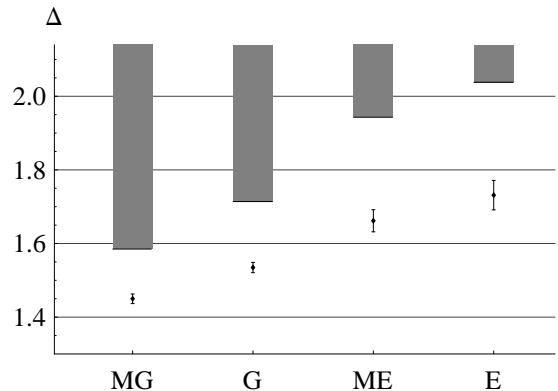}%
\caption{\label{res}Values of $\Delta$ for media in Table \protect\ref{ss}
(note reordering). Shaded bars: region excluded by replica bounds. Symbols:
numerical results (error bars indicate combined statistical and systematic
uncertainty).}
\end{figure}

The numerical results are plotted, along with the replica bounds of
Sec.~\ref{Explicit}, in Fig.~\ref{res}. Our numerical results are consistent
with, but substantially more precise than, the existing simulations described
in Sec.~\ref{Existing}. The replica bounds are seen to be not only valid but
also sharp within about 15\%. Furthermore, the relative order of $\Delta$ among
the media agrees with that of the replica bounds, supporting the validity of
the finite-band-renormalization picture discussed in Sec.~\ref{Replica}. The
significant variation of $\Delta$ among media, and the close agreement with
replica bounds, suggest that the replica formulas are useful and accurate also
for the practically important case of Huygens propagation in $d = 3$, although
no reliable data for comparison are known.

\section{\label{Discussion}Discussion}

Motivated by the problem of Huygens-front propagation in isotropic random
media, which reduces to previously studied white-noise systems (Burgers
turbulence and directed polymers) in the weak-perturbation limit, we have
performed a systematic study to obtain quantitative information about the front
speedup via the prefactor $\Delta$ of the already established $\epsilon^{4/3}$
scaling \cite{MK07}. The prefactor $\Delta$ corresponds to the energy density
of the Burgers fluid and the binding energy of the directed polymer, making
these ``toy models'' directly applicable to the propagation problem. The
latter, though also idealized, is physically more realistic because, e.g.,
white noise is not assumed. We have extended the variational analysis based on
the replica method---previously applied to directed polymers
\cite{MP91,G93,B94} and then to Burgers turbulence \cite{BMP95}---with a
specific focus on the value of $\Delta$ in the zero-temperature or inviscid
limit, corresponding to Huygens propagation. This analysis has been found
sufficiently tractable to yield explicit upper bounds on $\Delta$ for arbitrary
perturbation spectra $D(k)$, subject to the previously identified conditions
\cite{MK07} for reduction of the propagation problem to white noise.

Let us summarize how the numerical value of the replica bound on $\Delta$ can
be obtained for a $d$-dimensional random medium with a particular spectrum
$D(k)$, either specified analytically or determined from an experiment or
simulation to sufficient precision to perform the required computations. [A
turbulent energy spectrum $E(k)$ can be converted to an equivalent
quenched-medium spectrum using Eq.\ (\ref{EtoD}).] First we obtain the
functions $B_p(z)$ from Eq.\ (\ref{Bp}) for $p = 0$,~$1$, $2$,~$3$, and define
$\mu(z)$ by Eq.\ (\ref{muz}). A general replica bound formula is then Eq.\
(\ref{RBb}), where $z_* \ge 0$ is such that $\mu(z)$ is a decreasing function
for $0 < z < z_*$. If $\mu(z)$ is decreasing for all $z > 0$, we can take $z_*
= \infty$ and use a simpler formula, Eq.\ (\ref{RB}). An alternative bound
applicable in all cases, which may be better or worse (or even completely
uninformative), is the ``one-step'' result, Eq.\ (\ref{RB1}). Heuristically,
the one-step bound tends to dominate for single-scale media in high dimensions
or in which low wave numbers are suppressed (the two are related because the
``volume'' of low wave numbers has a $k^d$ factor).

The replica results are particularly interesting in light of rigorous
properties of random Huygens propagation that have been deduced from general
arguments. The dependence of the replica bounds on the form of the spectrum
(confirmed by numerical results) indicates that the rigorous bound on relative
speedup derived in Sec.~\ref{Spectrum} is not usefully sharp when applied to
spectra of different shapes. The replica bounds, at least for one class of
media, are consistent with the fact that a lower-dimensional ``slice'' through
a medium has a smaller (or equal) speedup due to elimination of some possible
paths. For randomly advected Huygens propagation (considered as an idealization
of turbulent combustion), monotonicity with respect to the laminar flame speed
$u_L$, in conjunction with the finite replica bound for weak $\mathrm{Re} =
\infty$ turbulence, precludes a divergent turbulent burning velocity $u_T$ in
strong $\mathrm{Re} = \infty$ turbulence ($u_L \to 0$).

The key qualitative insight obtained from the analytic form of the replica
bounds is the concept of finite-band renormalization. The picture of a
progressively coarse-grained medium with an upwardly renormalized propagation
speed has been previously used in turbulent combustion \cite{Y88,P94}, but it
was assumed that the renormalization is purely local in wave number, i.e., that
the effect on the renormalized speed $u_R$ of eliminating an arbitrarily narrow
high-wave-number band depends only on the spectral power $\phi$ in that band.
On dimensional grounds, then, the change in renormalized speed is
\begin{equation}
\label{duR}
\delta u_R \propto u_R \left(\frac{\phi}{u_R^2}\right)^{r/2},
\end{equation}
where we assume that the dependence on $\phi$ is a power law. If the spectrum
consisted only of the band in question, then Eq.\ (\ref{duR}) would have to
reproduce the weak-perturbation speedup scaling (now known to be $r =
\frac{4}{3}$). Equation (\ref{duR}) can be rewritten as a simple additive
renormalization,
\begin{equation}
\label{duRr}
\delta(u_R^r) \propto \phi^{r/2}.
\end{equation}
It was then argued \cite{P94} that in turbulent combustion, the effect of
eliminating many such bands in succession, covering an entire spectrum, would
be to increase $u_R$ from $u_L$ to $u_T$, with a cumulative renormalization of
$u_R^r$ proportional to $(u'^2)^{r/2}$ (where $u'^2$ is the total spectral
power), giving
\begin{equation}
\label{Poch}
u_T^r - u_L^r \propto u'^r.
\end{equation}
This formula has the encouraging feature that $u_T \propto u'$ for $u_L \to 0$,
as expected.

We observe, however, that the use of arbitrarily narrow wave-number bands is
incompatible with $r = \frac{4}{3}$. If the spectrum is divided into $M$ bands
of equal power $\phi = u'^2/M$, then the effect of $M$ renormalizations by Eq.\
(\ref{duRr}) is
\begin{equation}
u_T^r - u_L^r \propto M \left(\frac{u'^2}{M}\right)^{r/2},
\end{equation}
which has a finite limit as $M \to \infty$ only if $r = 2$. This exponent
value, corresponding to an $\epsilon^2$ weak-perturbation speedup, was in fact
suggested by an earlier field-theoretic renormalization analysis based on
infinitesimal wave-number bands \cite{Y88}, which is widely used as a model of
$u_T$. That analysis, besides having the wrong weak-turbulence scaling,
predicts that $u_T$ depends only on $u_L$ and $u'$ but not on the form of the
spectrum, in contradiction to the nonuniversality (spectrum dependence) seen in
our analytical and numerical results.

The unsuitability of infinitesimal bands for analyzing Huygens propagation is
seen not only formally but also physically. The rationale for stepwise
renormalization is that the front reaches a steady state with respect to
small-scale perturbations, thereby determining the effective speed of a
coarse-grained front that responds to larger-scale perturbations. This picture
is literally applicable if the perturbations exist on two widely separated
scales. For continuous spectra, such renormalization is approximately justified
if the bands are wide enough to give significant scale separation, but narrow
enough to be roughly monochromatic so that spectral shape is not an issue
within each band. Thus it is not surprising that the replica bounds involve the
power in an order-unity band, Eq.\ (\ref{z32B}).

Front-speed renormalization has here been placed on a sounder footing by means
of the replica method, but only in the weak-perturbation limit. It is tempting
to conjecture that a relation like Eq.\ (\ref{Poch}) with $r = \frac{4}{3}$ may
still hold beyond that limit, with $u_T^{4/3} - u_L^{4/3}$ given by a quantity
characterizing the random advection. This would imply that the weak-turbulence
speedup prefactor $\Delta$ can be used to determine the strong-turbulence value
of $u_T$ by taking $u_L \to 0$. Such a relation, however, is incompatible with
the expectation that $u_T$ in strong turbulence depends not only on the
spectrum (or two-point spatial correlation function), which completely
determines $\Delta$, but also on other flow properties including time
dependence and higher moments. Time dependence clearly can affect $u_T$
because, e.g., if the flow correlation time goes to zero at fixed $u'$ and
$u_L$, then the turbulent diffusivity vanishes and the effect of advection
disappears. Dependence on only the two-point spatial correlation function is an
asymptotic result of the central limit theorem for $u'/u_L \to 0$ and does not
apply beyond that regime \cite{MK07}. Thus a relation like Eq.\ (\ref{Poch})
can hold only for a restricted class of flows, if at all. More generally, it is
not yet clear whether the finite-band renormalization concept is useful beyond
the weak-perturbation limit.

Because the accuracy of the replica bounds on $\Delta$ is not rigorously
established in general, it is important to seek independent validation. To this
end, we have presented the results of high-precision numerical simulations of
the one-dimensional inviscid KPZ-Burgers equation with white-in-time forcing,
which corresponds to two-dimensional Huygens propagation. The numerical results
are within about 15\% of the replica bounds for four example media. Although
the replica method has been validated rigorously for spin glasses
\cite{G03,T06}, the present work is the only quantitative test of
directed-polymer replica bounds known to us.

High-precision simulations in a greater number of dimensions would be very
costly. Well-controlled experiments would be an alternative way to validate the
replica results for three-dimensional propagation. One speculative possibility
is based on reinterpreting the function $T_0$ appearing in the eikonal equation
(\ref{eik0}) as an electrostatic potential. If a heterogeneous ferroelectric
medium can be constructed in which the electric field $-\bm{\nabla}T_0$ at each
point has a frozen magnitude $\propto 1 + \sigma(\mathbf{x})$ but an
unconstrained direction, then one face of the medium can be grounded ($T_0 =
0$), which determines the electric-field direction throughout the medium, and
the potential can be measured along the opposite face to determine the
``speedup.'' (A small dissipative term $\propto \nabla^2 T_0$ is needed in the
eikonal equation to regulate singularities in accordance with Huygens'
principle, and so the local charge density must affect the mechanism that
freezes the local electric-field magnitude.) By whatever technique,
confirmation of the sharpness of the replica bounds for three-dimensional
propagation under idealized conditions would establish these bounds as an
appropriate starting point and limiting case for more complex and realistic
engineering models, such as are needed in turbulent combustion. Other physical
applications were noted previously \cite{MK07}.

\begin{acknowledgments}
The U.S. Department of Energy, Office of Basic Energy Sciences, Division of
Chemical Sciences, Geosciences, and Biosciences supported this work. Sandia is
a multiprogram laboratory operated by Sandia Corporation, a Lockheed Martin
Company, for the U.S. Department of Energy under contract DE-AC04-94AL85000.
\end{acknowledgments}

\appendix

\section{\label{appKPZ}Applicability of the KPZ equation}

In Sec.~\ref{Relations} we assumed that the KPZ equation (\ref{KPZ}) adequately
describes the propagation of an initially flat front in a quenched medium with
weak random fluctuations. Here, the justification of this assumption is
explained. An exact equation for Huygens propagation in a quenched medium is
the eikonal equation \cite{MK07}
\begin{equation}
\label{eik0}
|\bm{\nabla}T_0| = 1 + \sigma,
\end{equation}
where $T_0(\mathbf{x})$ is the arrival time at a point $\mathbf{x}$ and
$\sigma(\mathbf{x})$ is the refractive-index fluctuation. If we define
$h(x_\parallel,\mathbf{x}_\perp) = x_\parallel -
T_0(x_\parallel,\mathbf{x}_\perp)$ in accordance with Eq.\ (\ref{Txh}) and
assume that the overall propagation is in the $+x_\parallel$-direction, then
the eikonal equation can be written
\begin{equation}
\label{eik}
\frac{\partial h}{\partial x_\parallel} = 1 - \sqrt{(1 + \sigma)^2 -
|\bm{\nabla}_\perp h|^2}.
\end{equation}

In an initial interval of $x_\parallel$ where $|\bm{\nabla}_\perp h|^2$ is
small compared to typical values of $\sigma$, the right-hand side of Eq.\
(\ref{eik}) equals $-\sigma$ to leading order. Thus the tilt $\bm{\nabla}_\perp
h$, initially zero, grows with $x_\parallel$ at a rate proportional to the
amplitude of $\sigma$ (measured by the rms fluctuation $\epsilon \ll 1$),
executing a random walk as new, uncorrelated fluctuations are encountered. It
follows that $|\bm{\nabla}_\perp h|^2$ remains smaller than $\epsilon$ for at
least a distance of order $\epsilon^{-1}$. (This is a conservative estimate
because cusp formation, equivalent to discarding certain branches of a
multivalued eikonal solution, can and does eliminate relatively large tilts.)

But the rescaling of $x_\parallel$ performed in Sec.~\ref{White} shows that the
characteristic distance for front equilibration is of order $\epsilon^{-2/3}
\ll \epsilon^{-1}$. Thus, at a minimum, our approximation $|\bm{\nabla}_\perp
h|^2 \ll \epsilon$ remains valid well after a statistically steady state is
reached, and its validity is then assured forever. Nonetheless, the
contribution of $|\bm{\nabla}_\perp h|^2$ must be included in a useful reduced
equation for propagation, because this nonlinear term is responsible for
producing the steady state. The leading terms in Eq.\ (\ref{eik}) then give the
inviscid KPZ equation
\begin{equation}
\frac{\partial h}{\partial x_\parallel} = \tfrac{1}{2} |\bm{\nabla}_\perp h|^2
- \sigma;
\end{equation}
the omitted terms are negligible compared to $|\bm{\nabla}_\perp h|^2$. We
conclude that the formation and properties of the Huygens-propagation steady
state (for sufficiently small $\epsilon$) are accurately described by the KPZ
equation with a non-white-noise perturbation $-\sigma$ and with viscosity taken
to zero.

\section{\label{appRSB}Gaussian trial functions and replica symmetry breaking}

To allow the expectation value $\langle\psi|\mathcal{H}_n|\psi\rangle$ of the
$n$-particle Hamiltonian (\ref{Hn}) to be expressed analytically in $n$, we
adopt the usual isotropic Gaussian trial wave functions
\begin{equation}
\psi(\mathbf{y}_1,\dotsc,\mathbf{y}_n) \propto \exp\biggl(-\tfrac{1}{4}
\sum_{a,b=1}^n (Q^{-1})_{ab}\, \mathbf{y}_a \cdot \mathbf{y}_b\biggr),
\end{equation}
which obey $\langle\psi|\, \mathbf{y}_a \cdot \mathbf{y}_b\, |\psi\rangle = (d
- 1) Q_{ab}$, where $Q$ is a positive-definite matrix. For positive integer
$n$, the optimal choice of $Q$ to approximate the ground state would be
invariant under arbitrary permutations of the replicas, because this is a
symmetry of the Hamiltonian. For $n \to 0$, however, at least within this
variational approach, the permutation invariance is violated through
hierarchical replica symmetry breaking \cite{MPV87,MP91}. The hierarchical
matrix $Q$ is constructed as follows, starting from positive integer $n$. Given
integers $1 \equiv m_0 \le m_1 \le \dotsb \le m_K \le m_{K+1} \equiv n$ such
that each $m_i$ divides $m_{i+1}$, define $M_i$ as the $n \times n$
block-diagonal matrix consisting of submatrices of size $m_i \times m_i$ with
all entries $1$ (e.g., $M_0$ is the identity matrix). Then define
\begin{equation}
\label{Q}
Q = \sum_{i=0}^{K+1} b_i M_i,
\end{equation}
where the scalars $b_i$ have dimensions $\di L^2$. The $n$ particles are thus
divided into blocks, sub-blocks, etc., that are bound on different length
scales; this already suggests a connection to renormalization ideas.

Because all the $M_i$ are seen to commute, the eigenvalues of $Q$ are readily
found. Each all-$1$ submatrix of $M_i$ has one eigenvalue $m_i$ and $m_i - 1$
eigenvalues $0$, so $M_i$ has $n/m_i$ eigenvalues $m_i$ and $n - n/m_i$
eigenvalues $0$. For $0 \le i \le K$, the matrix $Q$ has $n/m_i - n/m_{i+1}$
eigenvectors that are in the null space of $M_j$ for $j > i$ but in the nonzero
eigenspace of $M_j$ for $j \le i$. Thus an eigenvalue of $Q$ with multiplicity
$n/m_i - n/m_{i+1}$ is
\begin{equation}
\label{Lambdai}
\Lambda_i = \sum_{j=0}^i b_j m_j.
\end{equation}
It is convenient to define piecewise constant functions on the real interval $1
< u \le n$:
\begin{align}
\Lambda(u) &= \Lambda_i & &(m_i < u \le m_{i+1}),\\
\label{zu}
z(u) &= 2 \sum_{j=0}^i b_j & &(m_i < u \le m_{i+1}),
\end{align}
so that
\begin{equation}
\label{Lambdaz}
\Lambda(u) = \tfrac{1}{2} u\, z(u) - \tfrac{1}{2} \int_1^u dv\, z(v).
\end{equation}
There is a single further eigenvalue of $Q$ given by Eq.\ (\ref{Lambdai}) with
$i = K + 1$, corresponding to the eigenvector $(1,\dotsc,1)$. This represents a
center-of-mass translation, and the eigenvalue $\Lambda_{K+1}$ should go to
infinity in the ground state (complete freedom of the center of mass).

We can interpret $z(m_i) = 2 \sum_{j=0}^{i-1} b_j$ as the variance of each
component of interparticle separation,
\begin{equation}
z(m_i) = \langle\psi|\, [\hat{\mathbf{n}} \cdot (\mathbf{y}_a -
\mathbf{y}_b)]^2\, |\psi\rangle = Q_{aa} + Q_{bb} - 2Q_{ab},
\end{equation}
for indices $a$ and $b$ that are in the same size-$m_i$ block but different
size-$m_{i-1}$ blocks. [We see this because, from Eq.\ (\ref{Q}), $Q_{aa} =
Q_{bb} = \sum_{j=0}^{K+1} b_j$ and $Q_{ab} = \sum_{j=i}^{K+1} b_j$.] Using the
Gaussian identity (\ref{expzeta}), this time with $\zeta = i\mathbf{k}_\perp
\cdot (\mathbf{y}_a - \mathbf{y}_b)$, it follows that
\begin{equation}
\langle\psi|\exp[i\mathbf{k}_\perp \cdot (\mathbf{y}_a - \mathbf{y}_b)]\,
|\psi\rangle = \exp[-\tfrac{1}{2} |\mathbf{k}_\perp|^2\, z(m_i)].
\end{equation}
Then, from Eq.\ (\ref{Vy}), we obtain
\begin{equation}
\label{Vyayb}
\begin{split}
&\langle\psi|\, V(|\mathbf{y}_a - \mathbf{y}_b|)\, |\psi\rangle\\
&\qquad = \int \frac{d^{d-1} \mathbf{k}_\perp}{(2\pi)^{d-1}}\,
\exp[-\tfrac{1}{2} |\mathbf{k}_\perp|^2\, z(m_i)]\, D(|\mathbf{k}_\perp|)\\
&\qquad \equiv B_0\bm{(}z(m_i)\bm{)},
\end{split}
\end{equation}
where the function $B_0$ is determined by the spectrum of the random medium.
The number of index pairs $(a,b)$ of the type considered is $n(m_i - m_{i-1})$;
the range $1 \le i \le K + 1$ covers all pairs with $a \neq b$. There are an
additional $n$ self-pairs ($a = b$) for which, in place of Eq.\ (\ref{Vyayb}),
we have $\langle\psi|\, V(0)\, |\psi\rangle = B_0(0)$. Thus
\begin{equation}
\begin{split}
&\sum_{a,b=1}^n \langle\psi|\, V(|\mathbf{y}_a - \mathbf{y}_b|)\,
|\psi\rangle\\
&\qquad = n\, B_0(0) + \sum_{i=1}^{K+1} n(m_i - m_{i-1})\,
B_0\bm{(}z(m_i)\bm{)}\\
&\qquad = n\left(B_0(0) + \int_1^n du\, B_0\bm{(}z(u)\bm{)}\right).
\end{split}
\end{equation}

Meanwhile, for the kinetic part of the Hamiltonian, we find
\begin{equation}
\begin{split}
\sum_{a=1}^n \langle\psi|\nabla_a^2|\psi\rangle &= -\tfrac{1}{4} (d - 1) \tr
Q^{-1}\\
&= -\tfrac{1}{4} (d - 1) \sum_{i=0}^K \left(\frac{n}{m_i} -
\frac{n}{m_{i+1}}\right) \frac{1}{\Lambda_i}\\
&= -\tfrac{1}{4} (d - 1)n \int_1^n \frac{du}{u^2\, \Lambda(u)},
\end{split}
\end{equation}
where we have set $1/\Lambda_{K+1} = 0$. Combining these results, we obtain the
expectation value of the Hamiltonian (\ref{Hn}),
\begin{equation}
\begin{split}
\langle\psi|\mathcal{H}_n|\psi\rangle = {}&\tfrac{1}{8} \tau^2 (d - 1)n
\int_1^n \frac{du}{u^2\, \Lambda(u)}\\
&- \frac{n}{2\tau} \left(B_0(0) + \int_1^n du\, B_0\bm{(}z(u)\bm{)}\right).
\end{split}
\end{equation}
This expression is very similar to the one obtained in a replica treatment of
directed polymers focused on the Gaussian medium \cite{B94}. There, however,
the factor $\frac{1}{2}$ multiplying the $B_0$ terms was incorrectly omitted.
Also, the self-interaction term $\propto B_0(0)$ was not shown (a constant
offset that does not affect the variational optimization but is important for
absolute energy values); the factor $1/\tau$ in Eq.\ (\ref{itse}) was included
in the definition of the Hamiltonian $\mathcal{H}_n$; and the following
notations were used: $N = d - 1$, $\beta = 1/\tau$, $\lambda(u) = [2\,
\Lambda(u)]^{-1}$, $Q(u) = z(u)$, $\hat f = -B_0/(d - 1)$.

In the formal limit $n \to 0$, the block ``sizes'' $m_i$ are assumed to be real
numbers in the reversed sequence $n \equiv m_{K+1} \le m_K \le \dotsb \le m_1
\le m_0 \equiv 1$, with arbitrarily large $K$ and no divisibility constraints.
As $K \to \infty$, then, $\Lambda(u)$ and $z(u)$ become general functions on
the interval $0 < u < 1$. Because $\Lambda(u)$ represents an eigenvalue of $Q$
and $z(u)$ represents a variance, both must be nonnegative. A further
constraint arises when we require nonnegativity of variances involving
arbitrary numbers of particles from various blocks (arbitrary because, upon
analytic continuation in $n$, there is no limit on the number of particle
indices that can be formally considered, unlike the case of positive integer
$n$). The parameters $b_i$, which determine the matrix elements of $Q$, must be
such that all eigenvalues $\Lambda(u)$ are nonnegative, not just for $n \to 0$
but also for arbitrary realizable integer values $1 \equiv m_0 \le \dotsb \le
m_{K+1} \equiv n$. From Eq.\ (\ref{Lambdai}), we see that nonnegativity of
$\Lambda_i$ for arbitrarily large $m_i$ requires $b_i \ge 0$. Consequently
$z(u)$, as defined in Eq.\ (\ref{zu}), is a nondecreasing function for $1 < u <
n$; and so in the $n \to 0$ limit, where the order of the $m_i$ is reversed,
$z(u)$ must be a \emph{nonincreasing} function for $0 < u < 1$. In fact, from
Eq.\ (\ref{Lambdaz}) and its implication
\begin{equation}
\label{Lpzp}
\Lambda'(u) = \tfrac{1}{2} u\, z'(u),
\end{equation}
a nonnegative and nonincreasing function $z(u)$ will produce a function
$\Lambda(u)$ with the same properties.

\section{\label{appNumerical}Numerical method for the inviscid KPZ equation}

We simulate the one-dimensional inviscid KPZ equation (\ref{KPZ1}) by a
Lagrangian finite-element method in which the only approximation, aside from
the periodic domain, is a time and space discretization of the white noise
$\eta$. Equation (\ref{KPZ1}) itself is solved exactly (except for roundoff
error). We solve this equation, rather than just the Burgers equation obtained
from it, because $h$ remains continuous at shocks and can be used to track
them, and because computing $h$ allows use of the formula $\Delta =
\langle\partial h/\partial t\rangle$ (in which averaging over $t$ is
particularly convenient). The time discretization of $\eta$ is in a sense the
simplest possible: a sequence of ``delta-function kicks'' at equally spaced
instants $t_k = kb$ with $k = 1$,~$2$, $\dotsc$. Each kick has a random
$x$~profile (of a form to be described) and an overall amplitude that scales
with $b^{1/2}$, to produce white noise as $b \to 0$.

\begin{figure*}
\includegraphics[width=178mm]{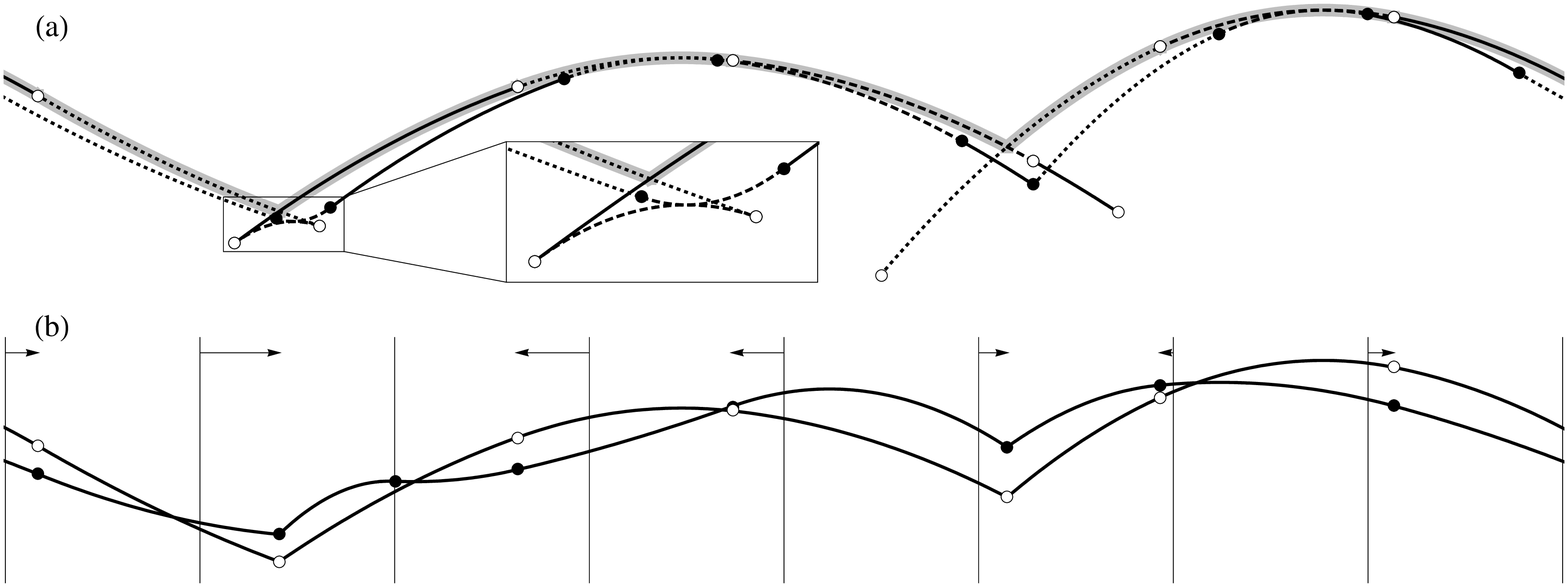}%
\caption{\label{front}Lagrangian finite-element method for the inviscid KPZ
equation. Plots show stages in the simulated evolution of $h(x)$ on a periodic
$x$ domain. (a)~Filled circles bound quadratic elements of the initial
configuration, distinguished by alternating dotted, dashed, and solid curves.
One initial cusp is present, in the right half of the plot. After advection of
all boundaries, $h(x)$ consists of corresponding elements bounded by open
circles, and is no longer single-valued. The filled circle at the initial cusp
separates into two open circles, and a dashed element in the left half of the
plot turns inside out when its boundaries pass through one another (inset). The
trimming procedure then constructs a single-valued $h(x)$ by retaining only the
largest values (shaded band). As a result, a solid element disappears into the
existing cusp, and the inside-out dashed element disappears to form a new cusp.
(b)~Open circles indicate trimmed element boundaries from (a), including
updated cusp locations. Vertical lines form grid for kicking. Arrows indicate
deformation of kicking boundaries to existing element boundaries, except third
grid line from left, which has no nearby element boundary and introduces a new
one. Kicking produces elements bounded by filled circles, without altering the
number or location of cusps. This final $h(x)$ can then be evolved again as
in~(a).}
\end{figure*}

The steps in our numerical method are displayed in Fig.~\ref{front}. We
represent a ``snapshot'' of $h(x)$ by a piecewise quadratic function on various
$x$~intervals (elements). Continuity of $h$ is required, but $\partial
h/\partial x$ can jump discontinuously upward at shocks (corresponding to
Huygens cusps that are concave, not convex). Between kicks, the elements evolve
dynamically in a way that preserves the piecewise quadratic form. Specifically,
the nonlinear term $\frac{1}{2} (\partial h/\partial x)^2$ in Eq.\ (\ref{KPZ1})
is quadratic in $x$ if $h$ is, and so when $\eta = 0$ the exact solution
remains in the piecewise quadratic space. Element boundaries without shocks are
simply advected at the local Burgers velocity $w = -\partial h/\partial x$
(which is constant along Lagrangian trajectories) \cite{S85}. For boundaries
with shocks, the adjacent segments are at first allowed to overlap. Even with
no shocks initially, elements can overlap if boundaries pass through one
another, indicating formation of a new shock.

In this way $h(x)$ can be evolved directly to the time of the next kick, but it
generally becomes multivalued, consisting of quadratic functions on overlapping
$x$~intervals. The solution is then ``trimmed'' to obtain nonoverlapping
elements containing the largest value of $h$ at each $x$, as illustrated in
Fig.~\ref{front}(a). Some elements are cut down; others are discarded entirely
(such as those that turn inside out when a new shock forms). The portions
removed can be interpreted as Burgers fluid elements that have run into shocks
or as segments of the Huygens front that have been overtaken at cusps. The
trimming procedure is tractable if no overlaps occur between elements that were
more distant than next-to-nearest neighbors. If this condition is violated, we
split the time interval in half and perform the evolution in two stages,
recursively. Because the configuration of $h(x)$ results from a random process,
the next-to-nearest interaction is in fact sufficient for evolution over a
finite time interval. That is, barring exact synchronization between different
parts of space (which is vanishingly unlikely), any overlap of more distant
elements can be reduced to discrete stages of evolution in which only nearest
and next-to-nearest neighbors overlap. For example, a shock gradually absorbs
the elements on either side of it (say elements $1$ and $2$), but one of the
two will disappear first, resulting in a renumbering of the remaining elements.
Only if they disappeared at the same instant would we have an unavoidable
interaction between elements $0$ and $3$.

``Kicking,'' illustrated in Fig.~\ref{front}(b), preserves the assumed form of
$h(x)$ if the spatial profile of the kick is also piecewise quadratic. Such a
profile is synthesized from a given spectrum $D(k)$ by first generating
delta-function spikes on a uniform $x$~grid (spacing $\delta$) based on a
spectrum $k^6\, D(k)$, and then forming a smooth piecewise quadratic function
by in effect analytically integrating three times with respect to $x$. (This
kick profile has an everywhere continuous derivative to avoid introducing
additional shocks, especially ones of unphysical sign.) The choice of the
forcing grid is the only way in which a finite spatial resolution $\delta$
enters the simulation. As described, the kicking process would introduce a new
element boundary in $h(x)$ at every grid point, since generically every
boundary introduced in a previous kick has moved at least slightly. Each
boundary survives for some characteristic timescale before vanishing into a
shock, and so the steady-state average number of boundaries present is
proportional to the rate at which they are introduced, which would scale with
the kicking rate $1/b$. As we take $b \to 0$ to represent white noise
accurately, we would have an explosion in the number of elements, but they
would be mostly redundant since the spatial resolution of the driving noise is
still limited by the grid. We adopt a more efficient approach that adds a new
boundary only when $h(x)$ is insufficiently resolved in the neighborhood of the
point in question, and otherwise deforms the kick so as to take advantage of an
existing nearby boundary rather than insisting on the planned grid. The
small-scale deformation distorts the high-wave-number part of the noise
spectrum and thus requires a somewhat finer grid to achieve the same accuracy.
But the cost is outweighed by the much-improved behavior as $b \to 0$: Now the
average number of boundaries remains fixed.

Conventional numerical methods for the Burgers-KPZ and similar equations
require fine spatial grids to resolve shocks, and nonzero viscosity to
stabilize them. By contrast, we take the inviscid limit from the start,
allowing an explicit geometric representation of shocks. The spatial grid need
only resolve the forcing; inviscid shocks are perfectly sharp and do not
introduce smaller length scales. If the forcing is spatially smooth (as for
media G and MG), then there is a significant advantage in efficiency from using
a relatively coarse grid and still capturing sharp shocks. With spatially rough
forcing (i.e., a spectrum with a long high-wave-number tail, as for media E and
ME), the advantage is less clear because a fine grid is needed anyway to
resolve the forcing accurately. Nevertheless, because we work directly at
$\mathrm{Re}_{\text{B}} = \infty$, there is one fewer parameter contributing
systematic errors. Our method is designed explicitly for a one-dimensional
simulation ($d = 2$); generalization to the Burgers-KPZ equation in two or more
dimensions ($d \ge 3$) is possible in principle, but the computational geometry
would be much more intricate and possibly intractable. As with conventional
methods, the computational cost of simulations in higher dimensions would be
severe.

We now describe how the parameters of our simulations are chosen and how the
uncertainties in the results are estimated. By systematic error we mean the
difference between the precise average of a quantity over many runs of a
practical simulation (where each run takes a finite computation time) and the
precise average for the idealized problem statement (where space and time are
considered infinite and continuous). An efficient computational approach
involves balancing statistical and systematic errors, both of which contribute
to the overall uncertainty of the result. In our case, the relevant systematic
parameters that ideally approach infinity are $L$, $1/\delta$, $1/b$, and the
time $T$ allowed for equilibration before data are taken. We estimate $\Delta$
by subsequently averaging $\partial h/\partial t$ over a time interval $3T$,
based on the following considerations: It cannot be efficient to spend much
more time on equilibration than on averaging, and even if it were ideal to
spend only a tiny fraction of the computation time on equilibration, spending
$\frac{1}{4}$ on it reduces the available data for averaging only modestly,
increasing the statistical error by $(\frac{4}{3})^{1/2} - 1 \simeq 15\%$.

Our framework for treating systematic errors is a conservative assumption that
these errors scale with the reciprocal of the  parameters given above. (This is
the slowest convergence that we would reasonably anticipate.) That is, we
assume that the precise average computed from many runs of a simulation tends
to the true value of $\Delta$ with asymptotic corrections of order $1/L$, order
$\delta$, order $b$, and order $1/T$. Consequently, we can extrapolate from
simulations performed with different finite parameters to estimate the result
of an ideal simulation. Because we do not trust this extrapolation as a
quantitative model, we will apply it only when all the simulation results
contributing to the extrapolation are statistically indistinguishable
(consistent with identical underlying averages).

Specifically, our final extrapolation will be based on a ``most refined''
simulation, with parameter set $\alpha$, and $N = 4$ lesser simulations
$\beta_0$, $\dotsc$, $\beta_{N-1}$, each based on $\alpha$ with one parameter
halved. All these simulations are repeated as necessary to obtain averages $A$
and $B_i$ with some statistical uncertainty $\sigma$ (to be determined below).
Because of the parameter halving and the assumed reciprocal scaling of
systematic errors, the extrapolation takes the simple form
\begin{equation}
\label{C2}
\Delta \simeq A + (A - B_0) + \dotsb + (A - B_{N-1}).
\end{equation}
To ensure that the result is fairly insensitive to our specific model of
systematic errors, we require each correction term $A - B_i$ to be within two
standard deviations of zero, i.e.,
\begin{equation}
\label{twosd}
\lvert A - B_i\rvert \le 2\sqrt2\, \sigma.
\end{equation}
(Taking a difference of two independent random variables multiplies the
standard deviation by $\sqrt2$.) If the decay of systematic errors is more
rapid than assumed, Eq.\ (\ref{C2}) is needlessly \emph{imprecise} but not
\emph{incorrect}, because the exaggerated correction terms are statistically
equivalent to zero. Our final estimate of $\Delta$, being a sum of independent
random variables $(N + 1)A - B_0 - \dotsb - B_{N-1}$, has a variance $[(N +
1)^2 + N]\sigma^2 = 29\sigma^2$. Thus $\sigma$ should be taken as the desired
overall uncertainty divided by $\sqrt{29}$. By repeated doubling, we can locate
a parameter set $\alpha$ sufficiently refined that Eq.\ (\ref{twosd}) holds.

The purpose of the extrapolation method is to obtain a conservative assessment
of the uncertainty contributed by systematic errors. Our procedure is
qualitatively similar to traditional rules of thumb, such as, ``The change in
the simulation result upon doubling the resolution is an estimate of the
systematic uncertainty.'' But the somewhat arbitrary technique of doubling the
resolution (or other parameters) is replaced here by a more fundamental
assumption about the scaling of systematic errors. (In simpler problems, such
as the finite-difference solution of deterministic differential equations, the
correct scaling can be readily obtained by analysis, but we desire a robust
``black box'' method.)

\begin{table}
\caption{\label{num}Most refined numerical simulation parameters, and resulting
extrapolated estimate of $\Delta$, for media in Table \protect\ref{ss} (with $a
= 1$).}
\begin{ruledtabular}
\begin{tabular}{llllll}
Medium &$\log_2 L$ &$\log_2(3\delta)$ &$\log_2 b$ &$\log_2 T$ &$\Delta$\\\hline
G  &$8$ &$-2$ &$-4$ &$8$ &$1.535(14)$\\
MG &$7$ &$-3$ &$-6$ &$6$ &$1.450(13)$\\
E  &$7$ &$-5$ &$-5$ &$5$ &$1.73(4)$\\
ME &$6$ &$-5$ &$-5$ &$4$ &$1.66(3)$\\
\end{tabular}
\end{ruledtabular}
\end{table}

For each of our four example spectra, Table \ref{num} gives the parameters
$\alpha$ of the most refined simulation performed, and the resulting
extrapolated estimate of $\Delta$ including both statistical and systematic
uncertainties.


\begin{thebibliography}{28}
\expandafter\ifx\csname natexlab\endcsname\relax\def\natexlab#1{#1}\fi
\expandafter\ifx\csname bibnamefont\endcsname\relax
  \def\bibnamefont#1{#1}\fi
\expandafter\ifx\csname bibfnamefont\endcsname\relax
  \def\bibfnamefont#1{#1}\fi
\expandafter\ifx\csname citenamefont\endcsname\relax
  \def\citenamefont#1{#1}\fi
\expandafter\ifx\csname url\endcsname\relax
  \def\url#1{\texttt{#1}}\fi
\expandafter\ifx\csname urlprefix\endcsname\relax\def\urlprefix{URL }\fi
\providecommand{\bibinfo}[2]{#2}
\providecommand{\eprint}[2][]{\url{#2}}

\bibitem[{\citenamefont{Kerstein and Ashurst}(1994)}]{KA94}
\bibinfo{author}{\bibfnamefont{A.~R.} \bibnamefont{Kerstein}} \bibnamefont{and}
  \bibinfo{author}{\bibfnamefont{W.~T.} \bibnamefont{Ashurst}},
  \bibinfo{journal}{Phys. Rev. E} \textbf{\bibinfo{volume}{50}},
  \bibinfo{pages}{1100} (\bibinfo{year}{1994}).

\bibitem[{\citenamefont{Mayo and Kerstein}(2007)}]{MK07}
\bibinfo{author}{\bibfnamefont{J.~R.} \bibnamefont{Mayo}} \bibnamefont{and}
  \bibinfo{author}{\bibfnamefont{A.~R.} \bibnamefont{Kerstein}},
  \bibinfo{journal}{Phys. Lett. A} \textbf{\bibinfo{volume}{372}},
  \bibinfo{pages}{5} (\bibinfo{year}{2007}).

\bibitem[{\citenamefont{Gomes et~al.}(2005)\citenamefont{Gomes, Iturriaga,
  Khanin, and Padilla}}]{GIKP05}
\bibinfo{author}{\bibfnamefont{D.}~\bibnamefont{Gomes}},
  \bibinfo{author}{\bibfnamefont{R.}~\bibnamefont{Iturriaga}},
  \bibinfo{author}{\bibfnamefont{K.}~\bibnamefont{Khanin}}, \bibnamefont{and}
  \bibinfo{author}{\bibfnamefont{P.}~\bibnamefont{Padilla}},
  \bibinfo{journal}{Moscow Math. J.} \textbf{\bibinfo{volume}{5}},
  \bibinfo{pages}{613} (\bibinfo{year}{2005}).

\bibitem[{\citenamefont{Bouchaud et~al.}(1995)\citenamefont{Bouchaud, M\'ezard,
  and Parisi}}]{BMP95}
\bibinfo{author}{\bibfnamefont{J.~P.} \bibnamefont{Bouchaud}},
  \bibinfo{author}{\bibfnamefont{M.}~\bibnamefont{M\'ezard}}, \bibnamefont{and}
  \bibinfo{author}{\bibfnamefont{G.}~\bibnamefont{Parisi}},
  \bibinfo{journal}{Phys. Rev. E} \textbf{\bibinfo{volume}{52}},
  \bibinfo{pages}{3656} (\bibinfo{year}{1995}).

\bibitem[{\citenamefont{M\'ezard and Parisi}(1991)}]{MP91}
\bibinfo{author}{\bibfnamefont{M.}~\bibnamefont{M\'ezard}} \bibnamefont{and}
  \bibinfo{author}{\bibfnamefont{G.}~\bibnamefont{Parisi}},
  \bibinfo{journal}{J. Phys. I (France)} \textbf{\bibinfo{volume}{1}},
  \bibinfo{pages}{809} (\bibinfo{year}{1991}).

\bibitem[{\citenamefont{Goldschmidt}(1993)}]{G93}
\bibinfo{author}{\bibfnamefont{Y.~Y.} \bibnamefont{Goldschmidt}},
  \bibinfo{journal}{Nucl. Phys. B} \textbf{\bibinfo{volume}{393}},
  \bibinfo{pages}{507} (\bibinfo{year}{1993}).

\bibitem[{\citenamefont{Blum}(1994)}]{B94}
\bibinfo{author}{\bibfnamefont{T.}~\bibnamefont{Blum}}, \bibinfo{journal}{J.
  Phys. A} \textbf{\bibinfo{volume}{27}}, \bibinfo{pages}{645}
  (\bibinfo{year}{1994}).

\bibitem[{\citenamefont{Kardar et~al.}(1986)\citenamefont{Kardar, Parisi, and
  Zhang}}]{KPZ86}
\bibinfo{author}{\bibfnamefont{M.}~\bibnamefont{Kardar}},
  \bibinfo{author}{\bibfnamefont{G.}~\bibnamefont{Parisi}}, \bibnamefont{and}
  \bibinfo{author}{\bibfnamefont{Y.-C.} \bibnamefont{Zhang}},
  \bibinfo{journal}{Phys. Rev. Lett.} \textbf{\bibinfo{volume}{56}},
  \bibinfo{pages}{889} (\bibinfo{year}{1986}).

\bibitem[{\citenamefont{Feynman and Hibbs}(1965)}]{FH65}
\bibinfo{author}{\bibfnamefont{R.~P.} \bibnamefont{Feynman}} \bibnamefont{and}
  \bibinfo{author}{\bibfnamefont{A.~R.} \bibnamefont{Hibbs}},
  \emph{\bibinfo{title}{Quantum Mechanics and Path Integrals}}
  (\bibinfo{publisher}{McGraw-Hill}, \bibinfo{address}{New York},
  \bibinfo{year}{1965}).

\bibitem[{\citenamefont{Williams}(1985)}]{W85}
\bibinfo{author}{\bibfnamefont{F.~A.} \bibnamefont{Williams}},
  \emph{\bibinfo{title}{Combustion Theory}}
  (\bibinfo{publisher}{Benjamin\slash\hspace{0pt}Cummings},
  \bibinfo{address}{Menlo Park, California}, \bibinfo{year}{1985}),
  \bibinfo{edition}{2nd} ed.

\bibitem[{\citenamefont{Crandall and Lions}(1983)}]{CL83}
\bibinfo{author}{\bibfnamefont{M.~G.} \bibnamefont{Crandall}} \bibnamefont{and}
  \bibinfo{author}{\bibfnamefont{P.-L.} \bibnamefont{Lions}},
  \bibinfo{journal}{Trans. Am. Math. Soc.} \textbf{\bibinfo{volume}{277}},
  \bibinfo{pages}{1} (\bibinfo{year}{1983}).

\bibitem[{\citenamefont{Sethian}(1985)}]{S85}
\bibinfo{author}{\bibfnamefont{J.~A.} \bibnamefont{Sethian}},
  \bibinfo{journal}{Commun. Math. Phys.} \textbf{\bibinfo{volume}{101}},
  \bibinfo{pages}{487} (\bibinfo{year}{1985}).

\bibitem[{\citenamefont{Roth et~al.}(1993)\citenamefont{Roth, M\"uller, and
  Snieder}}]{RMS93}
\bibinfo{author}{\bibfnamefont{M.}~\bibnamefont{Roth}},
  \bibinfo{author}{\bibfnamefont{G.}~\bibnamefont{M\"uller}}, \bibnamefont{and}
  \bibinfo{author}{\bibfnamefont{R.}~\bibnamefont{Snieder}},
  \bibinfo{journal}{Geophys. J. Int.} \textbf{\bibinfo{volume}{115}},
  \bibinfo{pages}{552} (\bibinfo{year}{1993}).

\bibitem[{\citenamefont{Kerstein and Ashurst}(1992)}]{KA92}
\bibinfo{author}{\bibfnamefont{A.~R.} \bibnamefont{Kerstein}} \bibnamefont{and}
  \bibinfo{author}{\bibfnamefont{W.~T.} \bibnamefont{Ashurst}},
  \bibinfo{journal}{Phys. Rev. Lett.} \textbf{\bibinfo{volume}{68}},
  \bibinfo{pages}{934} (\bibinfo{year}{1992}).

\bibitem[{\citenamefont{Iturriaga and Khanin}(2003)}]{IK03}
\bibinfo{author}{\bibfnamefont{R.}~\bibnamefont{Iturriaga}} \bibnamefont{and}
  \bibinfo{author}{\bibfnamefont{K.}~\bibnamefont{Khanin}},
  \bibinfo{journal}{Commun. Math. Phys.} \textbf{\bibinfo{volume}{232}},
  \bibinfo{pages}{377} (\bibinfo{year}{2003}).

\bibitem[{\citenamefont{Edwards and Anderson}(1975)}]{EA75}
\bibinfo{author}{\bibfnamefont{S.~F.} \bibnamefont{Edwards}} \bibnamefont{and}
  \bibinfo{author}{\bibfnamefont{P.~W.} \bibnamefont{Anderson}},
  \bibinfo{journal}{J. Phys. F} \textbf{\bibinfo{volume}{5}},
  \bibinfo{pages}{965} (\bibinfo{year}{1975}).

\bibitem[{\citenamefont{Binder and Young}(1986)}]{BY86}
\bibinfo{author}{\bibfnamefont{K.}~\bibnamefont{Binder}} \bibnamefont{and}
  \bibinfo{author}{\bibfnamefont{A.~P.} \bibnamefont{Young}},
  \bibinfo{journal}{Rev. Mod. Phys.} \textbf{\bibinfo{volume}{58}},
  \bibinfo{pages}{801} (\bibinfo{year}{1986}).

\bibitem[{\citenamefont{den Hollander and Toninelli}(2005)}]{DT05}
\bibinfo{author}{\bibfnamefont{F.}~\bibnamefont{den Hollander}}
  \bibnamefont{and}
  \bibinfo{author}{\bibfnamefont{F.}~\bibnamefont{Toninelli}},
  \bibinfo{journal}{Eur. Math. Soc. Newsl.} \textbf{\bibinfo{volume}{56}},
  \bibinfo{pages}{13} (\bibinfo{year}{2005}).

\bibitem[{\citenamefont{M\'ezard and Parisi}(1992)}]{MP92}
\bibinfo{author}{\bibfnamefont{M.}~\bibnamefont{M\'ezard}} \bibnamefont{and}
  \bibinfo{author}{\bibfnamefont{G.}~\bibnamefont{Parisi}},
  \bibinfo{journal}{J. Phys. A} \textbf{\bibinfo{volume}{25}},
  \bibinfo{pages}{4521} (\bibinfo{year}{1992}).

\bibitem[{\citenamefont{Guerra}(2003)}]{G03}
\bibinfo{author}{\bibfnamefont{F.}~\bibnamefont{Guerra}},
  \bibinfo{journal}{Commun. Math. Phys.} \textbf{\bibinfo{volume}{233}},
  \bibinfo{pages}{1} (\bibinfo{year}{2003}).

\bibitem[{\citenamefont{Talagrand}(2006)}]{T06}
\bibinfo{author}{\bibfnamefont{M.}~\bibnamefont{Talagrand}},
  \bibinfo{journal}{Ann. Math.} \textbf{\bibinfo{volume}{163}},
  \bibinfo{pages}{221} (\bibinfo{year}{2006}).

\bibitem[{\citenamefont{Sendi{\~n}a-Nadal
  et~al.}(1998)\citenamefont{Sendi{\~n}a-Nadal, Mu{\~n}uzuri, Vives,
  P\'erez-Mu{\~n}uzuri, Casademunt, Ram{\'\i}rez-Piscina, Sancho, and
  Sagu\'es}}]{SMVPCRSS98}
\bibinfo{author}{\bibfnamefont{I.}~\bibnamefont{Sendi{\~n}a-Nadal}},
  \bibinfo{author}{\bibfnamefont{A.~P.} \bibnamefont{Mu{\~n}uzuri}},
  \bibinfo{author}{\bibfnamefont{D.}~\bibnamefont{Vives}},
  \bibinfo{author}{\bibfnamefont{V.}~\bibnamefont{P\'erez-Mu{\~n}uzuri}},
  \bibinfo{author}{\bibfnamefont{J.}~\bibnamefont{Casademunt}},
  \bibinfo{author}{\bibfnamefont{L.}~\bibnamefont{Ram{\'\i}rez-Piscina}},
  \bibinfo{author}{\bibfnamefont{J.~M.} \bibnamefont{Sancho}},
  \bibnamefont{and} \bibinfo{author}{\bibfnamefont{F.}~\bibnamefont{Sagu\'es}},
  \bibinfo{journal}{Phys. Rev. Lett.} \textbf{\bibinfo{volume}{80}},
  \bibinfo{pages}{5437} (\bibinfo{year}{1998}).

\bibitem[{\citenamefont{Gotoh and Kraichnan}(1998)}]{GK98}
\bibinfo{author}{\bibfnamefont{T.}~\bibnamefont{Gotoh}} \bibnamefont{and}
  \bibinfo{author}{\bibfnamefont{R.~H.} \bibnamefont{Kraichnan}},
  \bibinfo{journal}{Phys. Fluids} \textbf{\bibinfo{volume}{10}},
  \bibinfo{pages}{2859} (\bibinfo{year}{1998}).

\bibitem[{\citenamefont{Gotoh}(1999)}]{G99}
\bibinfo{author}{\bibfnamefont{T.}~\bibnamefont{Gotoh}},
  \bibinfo{journal}{Phys. Fluids} \textbf{\bibinfo{volume}{11}},
  \bibinfo{pages}{2143} (\bibinfo{year}{1999}).

\bibitem[{\citenamefont{Gotoh}()}]{G06}
\bibinfo{author}{\bibfnamefont{T.}~\bibnamefont{Gotoh}}, \bibinfo{note}{private
  communication}.

\bibitem[{\citenamefont{Yakhot}(1988)}]{Y88}
\bibinfo{author}{\bibfnamefont{V.}~\bibnamefont{Yakhot}},
  \bibinfo{journal}{Combust. Sci. Tech.} \textbf{\bibinfo{volume}{60}},
  \bibinfo{pages}{191} (\bibinfo{year}{1988}).

\bibitem[{\citenamefont{Pocheau}(1994)}]{P94}
\bibinfo{author}{\bibfnamefont{A.}~\bibnamefont{Pocheau}},
  \bibinfo{journal}{Phys. Rev. E} \textbf{\bibinfo{volume}{49}},
  \bibinfo{pages}{1109} (\bibinfo{year}{1994}).

\bibitem[{\citenamefont{M\'ezard et~al.}(1987)\citenamefont{M\'ezard, Parisi,
  and Virasoro}}]{MPV87}
\bibinfo{author}{\bibfnamefont{M.}~\bibnamefont{M\'ezard}},
  \bibinfo{author}{\bibfnamefont{G.}~\bibnamefont{Parisi}}, \bibnamefont{and}
  \bibinfo{author}{\bibfnamefont{M.~A.} \bibnamefont{Virasoro}},
  \emph{\bibinfo{title}{Spin Glass Theory and Beyond}}
  (\bibinfo{publisher}{World Scientific}, \bibinfo{address}{Singapore},
  \bibinfo{year}{1987}).

\end{thebibliography}
\end{document}